\documentclass[aps,prx,twocolumn,nobibnotes,floatfix,superscriptaddress, nobalancelastpage]{revtex4-2}

\usepackage[utf8]{inp utenc}

\usepackage{graphicx,dblfloatfix}
\usepackage{bm}
\usepackage{amsmath}
\usepackage{amssymb}
\usepackage{braket} 
\usepackage{epsfig} 
\usepackage{tensor}
\usepackage{ifthen}
\usepackage{sidecap}
\usepackage{listings}
\usepackage[para,online,flushleft]{threeparttablex}

\usepackage{hyperref}
\usepackage[all]{hypcap}
\usepackage{float}

\renewcommand{\vec}[1]{\mbox{\boldmath $#1$}}

\usepackage{xcolor}
\definecolor{pastelgray}{rgb}{0.81, 0.81, 0.77}
\definecolor{beaublue}{rgb}{0.9, 0.9, 0.93}

\renewcommand{\vec}[1]{\mbox{\boldmath $#1$}}

\begin{document}

\title{Multiscale physics of atomic nuclei from first principles}

\author{Z.~H.~Sun}
\affiliation{Physics Division, Oak Ridge National Laboratory, Oak Ridge, Tennessee 37831, USA}

\author{A.~Ekstr\"om}
\affiliation{Department of Physics, Chalmers University of Technology, SE-412 96 G\"oteborg, Sweden}

\author{C. Forss{\'e}n}
\affiliation{Department of Physics, Chalmers University of Technology, SE-412 96 G\"oteborg, Sweden}

\author{G.~Hagen}
\affiliation{Physics Division, Oak Ridge National Laboratory, Oak Ridge, Tennessee 37831, USA}
\affiliation{Department of Physics and Astronomy, University of Tennessee, Knoxville, Tennessee 37996, USA}

\author{G.~R. Jansen}
\affiliation{National Center for Computational Sciences, Oak Ridge National Laboratory, Oak Ridge, TN 37831, USA}
\affiliation{Physics Division, Oak Ridge National Laboratory, Oak Ridge, Tennessee 37831, USA}

\author{T.~Papenbrock}
\affiliation{Department of Physics and Astronomy, University of Tennessee, Knoxville, Tennessee 37996, USA}
\affiliation{Physics Division, Oak Ridge National Laboratory, Oak Ridge, Tennessee 37831, USA}

\begin{abstract}
Atomic nuclei exhibit multiple energy scales ranging from hundreds of MeV in binding energies to fractions of an MeV for low-lying collective excitations. As the limits of nuclear binding is approached near the neutron- and proton driplines, traditional shell-structure starts to melt with an onset of deformation and an emergence of coexisting shapes. It is a long-standing challenge to describe this multiscale physics starting from nuclear forces with roots in quantum chromodynamics. Here we achieve this within a unified and non-perturbative quantum many-body framework that captures both short- and long-range correlations starting from modern nucleon-nucleon and three-nucleon forces from chiral effective field theory. The short-range (dynamic) correlations which accounts for the bulk of the binding energy is included within a symmetry-breaking framework, while long-range (static) correlations (and fine details about the collective structure) are included by employing  symmetry projection techniques. Our calculations accurately reproduce---within theoretical error bars---available experimental data for low-lying collective states and the electromagnetic quadrupole transitions in $^{20-30}$Ne. In addition, we reveal coexisting spherical and deformed shapes in $^{30}$Ne, which indicates the breakdown of the magic neutron number $N=20$ as the key nucleus $^{28}$O is approached, and we predict that the dripline nuclei $^{32,34}$Ne are strongly deformed and collective. By developing reduced-order-models for symmetry-projected states, we perform a global sensitivity analysis and find that the subleading singlet $S$-wave contact and a pion-nucleon coupling strongly impact nuclear deformation in chiral effective-field-theory. The techniques developed in this work clarify how microscopic nuclear forces generate the multiscale physics of nuclei spanning collective phenomena as well as short-range correlations and allow to capture emergent and dynamical phenomena in finite fermion systems such as atom clusters, molecules, and atomic nuclei. 
\end{abstract}

\maketitle


\section{Introduction}
\begin{figure}[!htbp]
\includegraphics[width=0.45\textwidth]{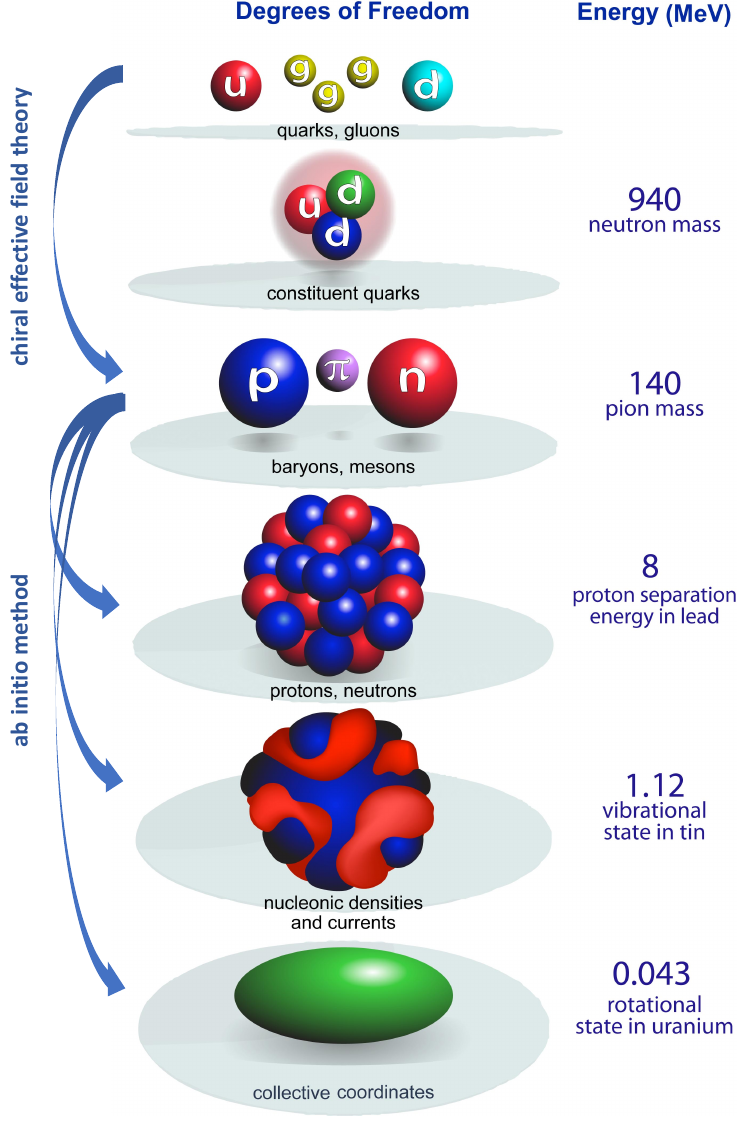}
\caption{{Energy scales and relevant degrees of freedom in nuclear physics, adapted from Ref.~\cite{LRP2007}. Also shown are the tools employed in this work: We use chiral effective field theory to relate interactions between nucleons to quantum chromodynamics and coupled-cluster theory as an ab initio method for the computation of binding energies and collective excitations at the highest resolution scale~\cite{Ekstrom:2022yea}.} }
\label{fig:fig1}
\end{figure}

%
Atomic nuclei exhibit emergent symmetry breaking: deformation allows rotation and is identified by strong electromagnetic transitions, i.e., large $B(E2)$ values, within states that belong to a rotational band~\cite{bohr1975}. While this has been established knowledge for more than 70 years, the multiscale description of such phenomena with inter-nucleon forces rooted in quantum chromodynamics has posed a long-standing challenge~\cite{wiringa2000,epelbaum2012,caprio2015,maris2015,jansen2016,togashi2016,yao2020,dytrych2020,miyagi2020,launey2020,caprio2021,Frosini:2021sxj,frosini2022,hagen2022,heller2022new}.  The situation  is illustrated in Fig.~\ref{fig:fig1}.
Binding energies, i.e., the negative of ground-state energies, are extensive quantities with about 8~MeV of binding per nucleon for medium-mass nuclei. The short range of the strong nuclear force implies that the bulk of the binding energy comes from short-range correlations in the wave function~\cite{coester1960}. Furthermore, as nucleons are fermions that interact via two- and three-nucleon forces, these short-range correlations are dominated by two-particle--two-hole and three-particle--three-hole excitations. Such dynamical correlations can be captured efficiently by various methods, and the numerical cost grows polynomially with increasing mass number~\cite{dickhoff2004,soma2013,hagen2014,hergert2016,stroberg2017}. 
The recent ab initio~\cite{Ekstrom:2022yea} computation
of the heavy nucleus $^{208}$Pb~\cite{hu2022}, for instance, is impressive because of the computational size of the problem. However, this doubly-magic nucleus has closed proton and neutron shells, and therefore exhibits a simple spherical structure. One ``only'' needs to capture dynamical correlations when computing its ground state. 
In contrast, open-shell nuclei are deformed and exhibit rotations. These introduce the lowest energy scale in atomic nuclei and range from hundreds of keV in medium-mass nuclei to tens of keV in heavy ones~\cite{bohr1975}, see also Fig.~\ref{fig:fig1}.  In the nucleus $^{34}$Mg, for instance, the lowest rotational excitation is only 0.26\% of the total binding energy~\cite{iwasaki2001}. A minuscule effect on the scale of nuclear energies that has a significant impact on the physics and shape of nuclei~\cite{Gaffney2013}. These collective modes involve long-range correlations that consist of many-particle--many-hole excitations. Such static correlations are difficult to capture. 

It is no surprise then that the computation of deformed nuclei is more challenging and so far limited to light and medium-mass nuclei~\cite{wiringa2000,epelbaum2012,caprio2015,maris2015,yao2020,dytrych2020,miyagi2020,caprio2021,Frosini:2021sxj,hagen2022}. 
The recent computations of deformed $p$-shell nuclei~\cite{caprio2015,maris2015,caprio2021} and neon nuclei~\cite{dytrych2020} underscore the importance of emergent symmetry breaking. However, these studies do not account for three-nucleon forces and---for neon---account for only a fraction of the  binding energy. The calculations by~\textcite{miyagi2020} include three-nucleon forces but did not reproduce the strong electric quadrupole transitions inside a rotational band. The calculations by~\textcite{yao2020} for $^{48}$Ti reproduce $B(E2)$ values but at the expense of somewhat too stretched energy spectra. The computations by~\textcite{Frosini:2021sxj} focus first on the static correlations and include dynamic correlations in a second step via perturbation theory. The latter accurately captures the binding energy only for sufficiently soft Hamiltonians (produced via a renormalization group transformation~\cite{bogner2007}) but that transformation changes the moment of inertia and thereby the collective properties. 

In this work we overcome these problems and demonstrate how to accurately capture multiscale physics of atomic nuclei starting from nuclear interactions rooted in quantum chromodynamics. We use the coupled-cluster~\cite{kuemmel1978,hagen2014} method for the non-perturbative inclusion of dynamical correlations and symmetry-projection techniques to capture static correlations. As we will see, this approach also correctly reproduces the electromagnetic transitions in a rotational band. 
We employ the accurate chiral interaction 1.8/2.0(EM) of Ref.~\cite{hebeler2011} which includes nucleon-nucleon forces at next-to-next-to-next-to leading-order (N3LO) and three-nucleon forces at next-to-next-to leading-order (NNLO). To better quantify our uncertainties in the predictions we also employ an ensemble of posterior interaction samples from chiral effective field theory with explicit delta degrees of freedom at NNLO. This ensemble was obtained in a recent study of $^{28}$O~\cite{kondo2023}. These interactions are described in detail in Sec.~\ref{sec:hamiltonian}. In order to capture deformation we start from an axially deformed Hartree-Fock reference state, and include short-range correlations using the coupled-cluster method ~\cite{kuemmel1978,bartlett2007,hagen2014}. In a final step we include long-range correlations by projecting the symmetry-broken coupled-cluster states on good angular momentum. Our projection is inspired by the disentangled approach by \textcite{qiu2017}, but avoids its shortcomings and extends it to electromagnetic transition matrix elements.

This work addresses the question ``What drives nuclear deformation?'', which has captivated generations of nuclear physicists~\cite{baranger1968,kumar1968,federman1979,nazarewicz1994,dufour1996,poves2018,tsunoda2020}. Let us briefly summarize some milestones in the description of nuclear deformation: In the 1950s, \textcite{bohr1952,bohr1953,nilsson1955} explained deformations as the surface vibrations of a liquid drop and the motion of independent nucleons confined inside~\cite{bohr1975}. In an alternative approach \textcite{elliott1958,elliott1958b} explained how deformed intrinsic states can be understood within the spherical shell model.
\textcite{dufour1996} revisited deformation in the nuclear shell model and found it useful to decompose the Hamiltonian into monopole and multipole parts~\cite{zuker1995,duflo1999}.
Here, the monopole essentially is the one-body normal-ordered term of the shell-model interaction, while the multipole terms are two-body operators; they contain the residual pairing and quadrupole interactions. 
These results have been succinctly summarized by Zuker's ``Multipole proposes, monopole disposes''\cite{poves2018}, i.e., the competition between pairing and quadrupole-quadrupole interactions might suggest deformation while the monopole---the effective spherical mean field---acts as a referee. We also note that the shell-model uses phenomenological ``effective charges'' to reproduce electric quadrupole transitions~\cite{caurier2005}. Nuclear density functional approaches confirmed the important role of the proton-neutron quadrupole-quadrupole interactions~\cite{dobaczewski1988,werner1994}; these approaches 
accurately describe deformation across the nuclear chart~\cite{stoitsov2003,delaroche2010,kortelainen2012,agbemava2014}. 

While we have a good understanding of nuclear deformation at low resolution scales, we lack insights how the high-resolution interactions from effective field theories of quantum chromodynamics cause it to emerge. While the pairing interaction can readily be identified with the nucleon-nucleon interaction in the $^1S_0$ partial wave, the origin of the quadrupole-quadrupole interaction is opaque at best or a pure shell-model concept at worst. With view on Ref.~\cite{federman1979}, one might be tempted to identify the quadrupole interaction with the isoscalar  $^3D_2$ partial wave (which is attractive). However, the quadrupole-quadrupole interaction is long range---in contrast to the short-range nucleon-nucleon interaction---and it is applicable only in model spaces consisting of one-to-two shells~\cite{baranger1968}. Thus, our understanding of nuclear deformation is still limited to a low-resolution shell-model picture. The ab initio computations~\cite{caprio2015,dytrych2020,takayuki2020,Frosini:2021sxj,hagen2022,becker2023} reproduced deformed nuclei but did not investigate how they are shaped by the underlying forces. In this work, we seek to understand what impacts deformation at the highest resolution scale possible today, i.e., based on chiral effective field theory
~\cite{epelbaum2009,machleidt2011,Hammer:2019poc}. To that aim we conduct a global sensitivity analysis~\cite{sobol2001} of collective nuclear properties and quantify how much individual terms in the chiral effective field theory interaction impact nuclear deformation. This global analysis is made possible using eigenvector continuation~\cite{frame2018}. Specifically, we develop a reduced-order model~\cite{duguet2023} for emulating~\cite{ekstrom2019,Konig:2019adq} ab initio calculations of deformed nuclei across millions of values for the low-energy constants in the chiral interaction. Our results are currently as close as we can get in tying low-energy nuclear structure to quantum chromodynamics without actually solving that non-Abelian gauge theory at low energies. 

Finally, we address a challenging problem regarding the theoretical description of shape coexistence in nuclei~\cite{heyde2011,gade2016,otsuka2020,nowacki2021,bonatsos2023}. The neutron-rich isotopes of neon and magnesium exhibit deformation and are a focus of experiments at rare isotope beam facilities~\cite{baumann2007,schwerdtfeger2009,doornenbal2009,wimmer2010,crawford2016,ahn2019,crawford2019,madurga2023}. In magnesium ($Z=12$) shape coexistence has been observed in $^{30}$Mg~\cite{schwerdtfeger2009} and $^{32}$Mg~\cite{wimmer2010,kitamura2021}, and the dripline is thought to be beyond $N=28$~\cite{baumann2007,crawford2019,tsunoda2020,macchiavelli2022}. The theoretical description of shape coexistence in $^{32}$Mg has been a challenge~\cite{reinhard1999,rodriguez2002,peru2014,caurier2014}.
The neon nuclei (proton number $Z=10$) are less known. No shape coexistence has been observed in $^{30}$Ne. The dripline nucleus is $^{34}$Ne~\cite{ahn2019}, and signatures of rigid rotation are found for $^{32}$Ne~\cite{doornenbal2009,murray2019}. The structure of $^{34}$Ne and the quadrupole transition strengths of $^{32,34}$Ne are unknown. This is a gap in our understanding in a critical region of the nuclear chart that extends towards the key nucleus $^{28}$O~\cite{kondo2023}.  For these reasons we focus on neutron-rich isotopes of neon for discovery and use neutron-rich magnesium nuclei for validation.

This paper is organized as follows. In Section~\ref{sec:theoframe} we describe the theoretical framework. We introduce  the Hamiltonian in Subsection~\ref{sec:hamiltonian}, describe the computation of reference states in Subsection~\ref{sec:no2b}, and for completeness we briefly review the coupled-cluster method in Subsection~\ref{sec:ccm}. Subsection~\ref{sec:projection} presents a novel approach to angular momentum projection within coupled-cluster theory applied to nuclei. In particular, this approach guarantees that norm and Hamiltonian kernels exhibit the correct symmetries. These developments might also be of interest for researchers in quantum chemistry. The computation of electromagnetic transition strengths is described in Subsection~\ref{sec:em}. In Subsection~\ref{sec:rom} we develop a reduced-order model for Hartree-Fock computations. This is a non-trivial extension of emulators based on eigenvector continuation because one has to ensure that only Slater determinants (and not superpositions thereof) enter.  We present our results in Sec.~\ref{sec:results} and summarize in Sec.~\ref{sec:summary}. A number of details and supporting material is presented in the appendices.  

\section{Theoretical framework}
\label{sec:theoframe}
\subsection{{Hamiltonian and model space.}}
\label{sec:hamiltonian}
We use the intrinsic Hamiltonian
\begin{equation}
H = T_{\rm kin} - T_\mathrm{CoM} + V_{NN} + V_{NNN}.
\label{Eq:hamiltonian} 
\end{equation}
Here $V_{NN}$ is the nucleon-nucleon ($NN$) potential, $V_{NNN}$ the three-nucleon ($NNN$) potential, $T_{\rm kin}$ the total kinetic energy, and $T_{\mathrm{CoM}}$ the kinetic energy of the center of mass. Using the intrinsic Hamiltonian effectively removes spurious center-of-mass motion~\cite{hagen2009a}. 

We employ various interactions in this work. For point predictions we use the 1.8/2.0(EM)~\cite{hebeler2011} interaction that yields accurate binding energies and spectra of light-, medium-, and heavy-mass nuclei~\cite{hagen2015,hagen2016b,simonis2017,morris2018,gysbers2019,hebeler2023}. It consists of an $NN$ potential at N3LO from Ref.~\cite{entem2003}, softened via similarity renormalization group transformation~\cite{bogner2007} to a momentum cutoff of 1.8 $\mathrm{fm}^{-1}$, and a bare $NNN$ potential at NNLO with a non-local regulator and a momentum cutoff of 2.0 $\mathrm{fm}^{-1}$. 

For posterior predictive distributions, incorporating relevant sources of uncertainty, we employ an ensemble of $n=100$ interactions that was calibrated in light-mass nuclei and recently used for accurate predictions of nuclei around  $^{28}$O~\cite{kondo2023}. These interactions are from chiral effective field theory at NNLO with explicit delta degrees of freedom. The $NN$ and $NNN$ potentials have non-local regulators and a momentum cutoff of 394~${\mathrm{MeV}}/c$~\cite{ekstrom2018,jiang2020}. This ensemble was obtained in Ref.~\cite{kondo2023} and consists of prior interaction samples filtered out by history matching~\cite{vernon2010,vernon2018,hu2022} to reproduce (within a non-implausibility window) scattering phase shifts, deuteron properties, the binding energies and charge radii of $^3$H, $^4$He, $^{16}$O, and ground- and excited states in $^{22,24,25}$O.
In order to use this ensemble for posterior predictions we proceed as follows: We first assign likelihood weights, $w_i = p(\mathcal{D}_\mathrm{cal} | \vec{\alpha}_i)$, with the excitation energies of the $J^\pi=2^+$ and $4^+$ rotational states in $^{24}$Ne as calibration data, $\mathcal{D}_\mathrm{cal}$, and $\vec{\alpha}_i$ a vector of  low-energy constants from the ensemble. For this, we use a simple normal likelihood that incorporates uncertainties from method, model space, and effective field theory truncations (see App.~\ref{app:error} for details). We then employ importance resampling~\cite{smith:1992aa,Jiang:2022off} with importance weights $q_i = w_i / \sum_{j=1}^n w_j$. This allows to effectively collect samples from the posterior
\begin{equation}
    p(\vec{\alpha}_i | \mathcal{D}_\mathrm{cal}) \propto p(\mathcal{D}_\mathrm{cal} | \vec{\alpha}_i) p(\vec{\alpha}_i) \ .
\end{equation}
Here we use a prior, $p(\vec{\alpha}_i)$, that is uniform for all low-energy constants except for $c_{1,2,3,4}$ where it corresponds to a Gaussian distribution from a Roy-Steiner analysis of pion-nucleon scattering~\cite{siemens2017}.
The posterior samples can then be used to make posterior predictions for rotational states and electromagnetic transitions in other nuclei (see Sec.~\ref{sec:neons}). There is a rather large fraction of 59 samples that have importance weight within one order of magnitude from the largest one, $q_\mathrm{max}$, and the effective number of samples is $n_\mathrm{eff} \equiv \sum_{i=1}^n q_i / q_\mathrm{max} = 25$.

\subsection{{Normal-ordered two-body approximation and computation of the reference state}}
\label{sec:no2b}

The inclusion of full $NNN$ forces in coupled-cluster computations is possible~\cite{hagen2007a} but  expensive. Fortunately, it is not necessary for accurate computations: Once a reference state is determined, one can employ the normal-ordered two-body approximation and discard residual three-nucleon terms from the Hamiltonian~\cite{hagen2007a,roth2012,binder2013}.

For open-shell nuclei, however,  the normal-ordered two-body Hamiltonian based on a deformed reference state breaks rotational symmetry. To avoid this problem we follow \textcite{Frosini:2021tuj} and first perform a spherical Hartree-Fock computation based on a uniform occupation of the partially filled shells. The resulting spherical density matrix is then used to make the normal-ordered two-body approximation.  The resulting normal-ordered two-body Hamiltonian is finally transformed back to the harmonic oscillator basis. This spherical two-body Hamiltonian is the starting point for our axially symmetric Hartree-Fock computation which then yields the deformed reference state $\vert \Phi_0 \rangle$. 

Our Hartree-Fock computations use a spherical harmonic oscillator basis of up to thirteen major shells while the $NNN$ interaction is further restricted by an energy cut $E_{\mathrm{3max}} = 16 \hbar\omega$. To gauge the convergence of our results we varied the harmonic oscillator frequency ($\hbar\omega$) from 10--16~MeV. Due to their computational cost, our angular-momentum projected coupled-cluster calculations are restricted to 8--9 major shells. This is sufficient to obtain spectra and quadrupole transitions that are converged with respect to the size of the model space for the nuclei we computed (see App.~\ref{app:benchmark} for details).

The computation of deformed reference states gives us the flexibility to study shape coexistence by targeting different deformations. The simplest approach is to fill the open shells according to the Nilsson model~\cite{nilsson1955} when initilaizing the density matrix for the ensuing Hartree-Fock computation. This allows one to construct prolate or oblate references. When computing nuclei with the ``magic'' neutron number $N=20$, this usually leads to reference states with a small deformation. Strongly deformed references can be obtained by adding the quadrupole constraint $\lambda r^2 Y_{20}(\hat{\mathbf{r}})$ to the Hamiltonian and by varying the Lagrange multiplier $\lambda$ such that a local energy minimum results (as a function of the quadrupole expectation value)~\cite{staszscak2010}. Thus, this is an important tool to study shape coexistence along the $N=20$ line for neutron-rich nuclei.

\subsection{{Coupled-cluster calculations include dynamical correlations}}
\label{sec:ccm}
Our coupled-cluster computations~\cite{kuemmel1978,bartlett2007,hagen2014}, \cite{bishop1991} start from an axially symmetric Hartree-Fock reference state $\vert \Phi_0 \rangle$ with prolate deformation~\cite{novario2020,hagen2022}. For a nucleus with mass number $A$, the coupled-cluster method parameterizes the many-nucleon wave-function as $\vert \Psi \rangle = e^T \vert \Phi_0 \rangle$, with $T = T_1 + T_2 + \ldots + T_A  $ being an expansion in $n$-particle--$n$-hole ($np$--$nh$) excitations ($n = 1, \ldots ,A$). To compute observables and transitions consistently we use the bi-variational coupled-cluster energy functional~\cite{arponen1982,arponen1983} where the left coupled-cluster state is parameterized as $\langle \widetilde{\Psi} \vert = \langle \Phi_0 \vert (1+\Lambda) e^{-T}$ with $\Lambda$ containing up to $np$--$nh$ de-excitations and truncated at the same order as $T$. For systems with a well-defined Fermi surface, the dynamical correlations---accounting for the bulk of the binding energy---are effectively captured by truncating $T \approx T_1 + T_2$, known as the coupled-cluster singles-and-doubles approximation (CCSD), and including $T_3$ perturbatively~\cite{bartlett2007,hagen2014}. This results in a polynomial scaling of computational cost of the order $N^6$ (or at most $N^7$) where $N$ is a measure of the system size~\cite{bartlett2007}. In both quantum chemistry and nuclear physics applications, CCSD is found to capture about 90\% of the full correlation energy. The inclusion of triples corrections brings that number up to about 99\%~\cite{bartlett2007,hagen2009b,hagen2014}. In App.~\ref{app:benchmark} we show that this is also the case for the deformed $^{20-34}$Ne isotopes using the 1.8/2.0(EM) chiral interaction. 


\subsection{{Angular momentum projection captures static correlations}}
\label{sec:projection}
We perform angular-momentum projections of deformed states computed in the CCSD approximation. For a more accurate angular-momentum projection than in Ref.~\cite{hagen2022} we use the bi-variational coupled-cluster energy functional~\cite{arponen1982,arponen1983} 
\begin{equation}
E^{(J)} = \frac{\langle \widetilde{\Psi} \vert P_J H \vert \Psi \rangle} {\langle \widetilde{\Psi} \vert P_J \vert \Psi \rangle }. 
\label{eq:CC_PAV}
\end{equation}
Here, $\vert \Psi \rangle \equiv e^T \vert \Phi_0 \rangle$ is the right coupled-cluster state and $\langle \widetilde{\Psi} \vert \equiv \langle \Phi_0 \vert (1 + \Lambda)e^{-T} $ is the corresponding left ground-state. $P_J$ is the angular-momentum projection operator 
\begin{equation}
\label{eq:PJ}
    P_{J} = \frac{2J+1}{2}\int\limits_0^\pi d\beta   \, d^J_{00}(\beta) R(\beta)\ ,
\end{equation}
and $R(\beta) = e^{-i\beta J_y}$ is the rotation operator.  
$P_J$ projects an axially symmetric state with $J_z=0$ onto a state with angular momentum $J$. This operator employs  the ``small'' Wigner $d_{00}^J(\beta)$ function, and $J_y$ is the $y$ component of the total angular momentum. To evaluate equation~(\ref{eq:CC_PAV}) we use the CCSD approximation and build on the disentangled approach by \textcite{qiu2017}. This approach applies the Thouless theorem~\cite{thouless1960} to act with the rotation operator $R(\beta)$ on the symmetry broken reference state, i.e., $\langle \Phi_0 \vert R(\beta) = \langle \Phi_0 \vert R(\beta) \vert \Phi_0 \rangle \langle \Phi_0 \vert e^{V(\beta)}$, with $V(\beta)$ being a $1p$--$1h$ de-excitation operator. Next, one expands 
\begin{equation}
\label{diff}
e^{V(\beta)}e^T=e^{W_0(\beta)+W_1(\beta)+W_2(\beta)+\ldots} \ .
\end{equation}
The series of $np$--$nh$ excitation operators $W_n$ does not truncate (even for $T \approx T_1 + T_2$) and includes up to $Ap$--$Ah$ excitations. 

\begin{figure}[h!]
    \includegraphics[width=0.99\linewidth]{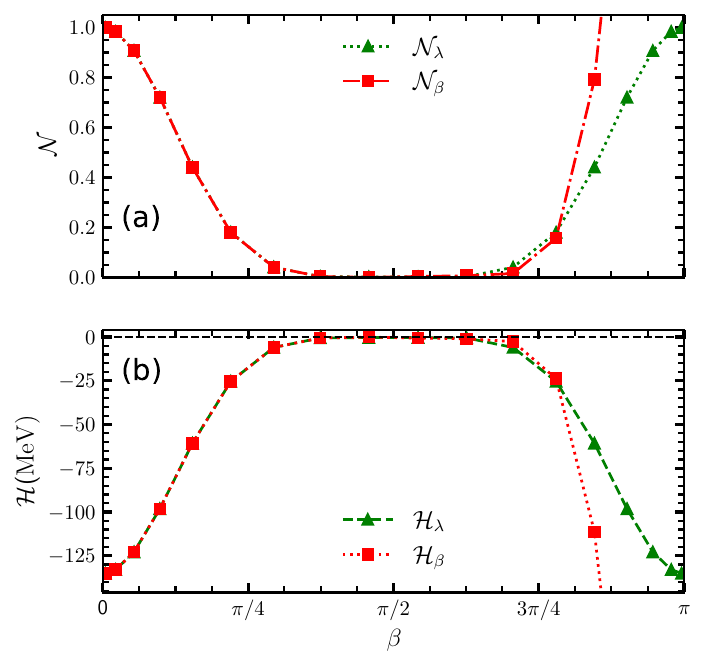}
    \caption{
     The upper panel (a) shows the norm kernels for $^{20}$Ne computed in the traditional disentangled approach (red, dash-dotted line) and with the new method of this work (green, dashed line). The lower panel (b) shows the same as (a) but for the Hamiltonian kernel. All calculations used the NNLO$_{\rm{opt}}$ interaction~\cite{ekstrom2013} with $\hbar\omega$=14 MeV in a model space of $N_{\rm{max}}=8$. The kernels  ($\mathcal{H}_\beta$, $\mathcal{N}_\beta)$ are the solutions of the differential equation when taking the derivative of Eq.~(\ref{diff}) with respect to $\beta$, while ($\mathcal{H}_\lambda$, $\mathcal{N}_\lambda$) are the solutions of the differential equation when taking the derivative of Eq.~(\ref{lambda_ode}) with respect to $\lambda$.}
    \label{fig:ne20_kernels}
\end{figure}

In this work we only keep the amplitudes $W_0, W_1$, and $W_2$. \textcite{qiu2017} proposed to compute the amplitudes $W_0, W_1$, and $W_2$ by taking the derivative of Eq.~(\ref{diff}) with respect to $\beta$. This leads to a set of ordinary differential equations with the initial $(\beta=0)$ values $W_0(0)=0, W_1(0)=T_1$, and $W_2(0)=T_2$. This approach has two disadvantages: First, $\frac{dV(\beta)}{d\beta}$ can have very large matrix elements in cases where the rotated state has a small overlap with the reference state, i.e. for $\langle \Phi_0 | R(\beta) |\Phi_0\rangle \approx 0$. This leads to a ``stiffness'' in the set of ordinary differential equations. Second, as one integrates the differential equations starting at $\beta=0$ the truncation at $W_2$ may lead to a loss of accuracy for larger angles. This loss of accuracy manifests itself, for instance, in a lack of symmetry of norm and Hamiltonian kernels under $\beta\to\pi-\beta$ for even-even nuclei, see Fig.~\ref{fig:ne20_kernels} for a numerical demonstration for the case of $^{20}$Ne.

To avoid these problems, we propose a new method to solve Eq.~(\ref{diff}) where we instead consider the expression
\begin{equation}\label{lambda_ode}
    e^{\lambda V}e^{T}=e^{W_0(\lambda)+W_1(\lambda)+W_2(\lambda)+\ldots} \ ,
\end{equation}
which, for $\lambda=1$, agrees with the previous one. Taking the derivative of Eq.~(\ref{lambda_ode}) with respect to $\lambda$ at fixed angle $\beta$ yields a new set of ordinary differential equations. We integrate over $\lambda$ from $0$ to $1$, and the initial values are $W_n(\lambda=0)=T_n$. These equations are solved at fixed $\beta$. Note that we suppressed the dependence of $V$ and $W_n$ on $\beta$ in equation~(\ref{lambda_ode}). This approach significantly improves the stability  of the numerical integration, keeps kernels symmetric under $\beta\to\pi-\beta$, and yields more accurate results for larger angles $\beta$. Results are shown in Fig.~\ref{fig:ne20_kernels}.
Nevertheless, the truncation at $W_2$ implies that the projection operator $P_J$ is not treated exactly, and angular momentum is only approximately a good quantum number.

\subsection{{Electromagnetic transition strengths}}
\label{sec:em}

The accurate computation of electromagnetic transitions strength in nuclei using ab initio methods has been a long-standing challenge~\cite{parzuchowski2017,HENDERSON2018468,miyagi2020,stroberg2022}. In this work we overcome this challenge by using symmetry projection techniques to capture the fine details in the nuclear wave-function that drives quadrupole collectivity. 

The electric quadrupole ($E2$) transition strength 
\begin{equation}
    B(E2,\downarrow) \equiv \vert\langle 0^+||Q_{2}||2^+\rangle\vert^2,
\end{equation}
is determined by the reduced matrix element $\langle 0^+||Q_{2}||2^+\rangle = \langle 0^+, J_z=0||Q_{20}||2^+,J_z=0\rangle/C^{00}_{2020}$. Here, $Q_{20}=\sum_j e(1/2-\tau_z^{(j)}r_j^2Y_{20}(\hat{\mathbf{r}}_j)$ is the electric quadrupole operator (given in terms of the electric charge $e$, the isospin operators $\tau_z^{(j)}$ and positions $\mathbf{r}_j$ of the nucleon labelled by $j$, and the spherical harmonics $Y_{20}$), and $C^{00}_{2020}$ is a Clebsch-Gordan coefficient. 
 
As coupled-cluster theory is based on a bi-variational functional, we need to compute  
\begin{align}
B(E2,\downarrow) & =  \frac{\langle \widetilde{\Psi} \vert P_{0} Q_{20} P_{2}\vert \Psi \rangle \langle \widetilde{\Psi} \vert P_{2} Q_{20} P_{0}\vert \Psi \rangle} {\langle \widetilde{\Psi} \vert P_0 \vert \Psi \rangle \langle \widetilde{\Psi} \vert P_2 \vert \Psi \rangle} 
\nonumber
\\
 & = \frac{\langle \widetilde{\Psi} \vert P_{0} Q_{20} \vert \Psi \rangle \langle \widetilde{\Psi} \vert Q_{20} P_{0}\vert \Psi \rangle} {\langle \widetilde{\Psi} \vert P_0 \vert \Psi \rangle \langle \widetilde{\Psi} \vert P_2 \vert \Psi \rangle}, 
 \label{eq:be2}
\end{align} 
where we removed redundant $P_2$ operators in the last step~\cite{bally2021}. We recall that $\vert \Psi \rangle \equiv e^T \vert \Phi_0 \rangle$ is the right coupled-cluster state and $\langle \widetilde{\Psi} \vert \equiv \langle \Phi_0 \vert (1 + \Lambda)e^{-T} $ is the corresponding left ground-state. Let us consider the computation of the matrix elements entering the numerator of equation~(\ref{eq:be2}).
The projector is based on the rotation operator $R$ (here we suppress the dependence on $\beta$), and we have
\begin{align}
    \langle \widetilde{\Psi} \vert R Q_{20} \vert \Psi \rangle &= \langle \Phi_0 \vert R \vert\Phi_0\rangle\langle \Phi_0\vert \widetilde{Z} \overline{Q}_{20} e^{W_0+W_1+W_2} \vert \Phi_0 \rangle,
\end{align}
with 
\begin{align}
\widetilde{Z} & = e^V R^{-1}(1+\Lambda)e^{-T}R e^{-V}, \\
\overline{Q}_{20} & = e^{V}{Q}_{20} e^{-V}.
\end{align}
Note that $\widetilde{Z}$ contains up to $2p$--$2h$ de-excitations, and that $\overline{Q}_{20}$ is a one-body similarity transformed operator.
The second part of the transition matrix element is 
\begin{align}
\nonumber
    \langle \widetilde{\Psi} \vert Q_{20} R \vert \Psi \rangle &= \langle \widetilde{\Psi} \vert R R^{-1} {Q}_{20} R  \vert \Psi  \rangle \\ 
    &= \langle \Phi_0 \vert R \vert\Phi_0\rangle\langle \Phi_0\vert \widetilde{Z} \widetilde{Q}_{20} e^{W_0+W_1+W_2} \vert \Phi_0 \rangle,
\end{align}
with 
\begin{align}
\widetilde{Q}_{20} & = e^V R^{-1}Q_{20} R e^{-V}. 
\end{align}
In App.~\ref{app:benchmark} we benchmark our approach with the symmetry-adapted no-core shell-model~\cite{launey2020} in $^{20}$Ne using the nucleon-nucleon potential $\mathrm{NNLO}_{\mathrm{opt}}$~\cite{ekstrom2013} and find agreement within estimated uncertainties.

\subsection{{A reduced-order model for projection-after-variation Hartree-Fock.}} 
\label{sec:rom}

We utilize eigenvector continuation~\cite{frame2018} to construct a reduced-order model~\cite{duguet2023} of Hartree-Fock. This enables us to develop fast and accurate emulators~\cite{Konig:2019adq,ekstrom2019} necessary for performing the sensitivity analyses presented in Sec.~\ref{sec:gsa}. We begin by exploiting that the delta-full NNLO Hamiltonian with $NN$ and $NNN$ interactions is a sum of terms with a linear dependence on the 17 low-energy constants $(\vec{\alpha})$ of interest, i.e.,
\begin{equation}
H(\vec{\alpha}) = H_{0} + \sum_{i=1}^{17}  \alpha_i H_i \   .
\label{eq:full_hamiltonian}
\end{equation}
Here, $H_i$ denote the respective Hamiltonian terms, and $H_0=T_{\rm{kin}}+V_0$ is the intrinsic kinetic energy $T_{\rm{kin}}$ and $V_0$ denotes $\vec{\alpha}$-independent  potential contributions such as one-pion exchange, leading two-pion exchange, and the Fujita-Miyazawa~\cite{fujita1957} $NNN$ interaction.

Let us consider a Hartree-Fock state $|\phi_i\rangle \equiv |\phi(\vec{\alpha}_i)\rangle$, with corresponding energy  $E_{\rm{HF}}(\vec{\alpha}_i)$ for some vector of values $\vec{\alpha}_i$. Clearly, the corresponding Hartree-Fock Hamiltonian is a one-body operator $H_{\rm{HF}}(\vec{\alpha}_i)$ that fulfills
\begin{equation}
    H_{\rm{HF}}(\vec{\alpha}_i)|\phi_i\rangle = E_{\rm{HF}}(\vec{\alpha}_i)|\phi_i\rangle \ .
\end{equation}
In general, given any Slater determinant $|\phi\rangle$ we can normal-order the Hamiltonian~(\ref{eq:full_hamiltonian}) with respect to $|\phi\rangle$ and obtain 
\begin{equation}
    H(\vec{\alpha}) = E_\phi(\vec{\alpha}) + F_\phi(\vec{\alpha}) +W_\phi(\vec{\alpha}) \ , 
\end{equation}
where $E_\phi(\vec{\alpha}) = \langle\phi|H(\vec{\alpha})|\phi\rangle$, and $F_\phi(\vec{\alpha})$ is the normal-ordered one-body Fock operator, and $W_\phi(\vec{\alpha})$ denotes any remaining terms. We have in particular 
\begin{equation}
\label{eq:HF_ham_def}
    H_{\rm HF}(\vec{\alpha}_i) = E_{\phi_i}(\vec{\alpha}_i) +F_{\phi_i}(\vec{\alpha}_i) \ .
\end{equation} 

We now seek to emulate the exact Hartree-Fock energy $E_\text{HF}(\vec{\alpha}_\circledcirc)$ for some target value $\vec{\alpha} = \vec{\alpha}_{\circledcirc}$ of the low-energy constants. To that end, we use eigenvector continuation~\cite{frame2018} and expand the target wave-function in a small set of Hartree-Fock states $\vert \phi_i\rangle$, i.e., a so-called snapshot basis, such that 
\begin{equation}
\vert \phi_{\circledcirc} \rangle \approx \sum_{i=1}^{N_\text{train}} c_i \vert \phi_i\rangle.
\end{equation}
The snapshot basis spans a low-dimensional subspace into which the Hartree-Fock Hamiltonian can be projected, thereby achieving a model-order reduction. We use the decomposition in Eq.~(\ref{eq:full_hamiltonian}) to project the individual interaction terms of the Hartree-Fock Hamiltonian to the subspace independently of $\vec{\alpha}$. This enables fast and accurate emulation of the Hartree-Fock energy $E_\text{HF}(\vec{\alpha}_\circledcirc)$ in the subspace for any target value $\vec{\alpha}_{\circledcirc}$ by solving the generalized eigenvalue problem 
\begin{equation} 
\label{eq:general_eigenvalue}
\sum_{ij}\langle {\phi}_i \vert H_{\rm{HF}}(\vec{\alpha}_{\circledcirc}) \vert \phi_j \rangle c_j = 
E_{\circledcirc} \sum_{ij} \langle {\phi}_i \vert \phi_j \rangle c_j.
\end{equation}
In our applications, see Sec.~\ref{sec:gsa}, we find $E_{\circledcirc} \approx E_\text{HF}(\vec{\alpha}_\circledcirc)$ with very high accuracy and precision using a very small basis of $N_\text{train}=68$ snapshots. 

It is important to recognize that the exact states, i.e., the snapshots, must be product states. Thus, one must not  replace $H_{\rm{HF}}$ in equation~(\ref{eq:general_eigenvalue}) with the full Hamiltonian, because this would correspond to the generator coordinate method (GCM)~\cite{hill1953,griffin1957} with the low-energy constants $\vec{\alpha}$ as continuous parameters. In the GCM case one would not reproduce the Hartree-Fock snapshots for $\vec{\alpha}_{\circledcirc} = \vec{\alpha}_i$, but rather obtain states corresponding to a lower energy from superpositions of product states.

To construct the Hartree-Fock Hamiltonian $H_{\rm{HF}}(\vec{\alpha}_{\circledcirc})$ in the subspace spanned by the snapshot basis we proceed as follows. We write the 
Hartree-Fock solution $\vert {\phi}_i \rangle $ for the snapshot value $\vec{\alpha}_i$ as
$\vert \phi_i \rangle = U_i \vert \Phi_0 \rangle$ where $\vert \Phi_0 \rangle$ is a reference state in the underlying harmonic-oscillator basis. The norm and Hamiltonian kernels for the generalized eigenvalue problem in the subspace~(\ref{eq:general_eigenvalue}) are  
\begin{eqnarray}
\label{eq:norm}
 \langle {\phi}_i \vert \phi_j \rangle  & = &  \langle \Phi_0 \vert \mathcal{O}_{ij} \vert \Phi_0\rangle \ , 
 \\
 \label{eq:hamiltonian}
   \langle {\phi}_i \vert H_{\rm{HF}}(\vec{\alpha}_{\circledcirc}) \vert \phi_j \rangle  & = & 
   \langle \Phi_0  \vert \mathcal{O}_{ij} h_j(\vec{\alpha}_{\circledcirc}) \vert \Phi_0 \rangle \ .
\end{eqnarray}
Here $\mathcal{O}_{ij} = U_i^\dagger U_j $ is a unitary matrix, and $h_j(\vec{\alpha}_{\circledcirc}) = U_j^\dagger H_{\rm{HF}}(\vec{\alpha}_{\circledcirc}) U_j$. With view on Eq.~(\ref{eq:HF_ham_def}) we now define
\begin{equation}
\label{eq:HF_gen_def}
    H_{\rm HF}(\vec{\alpha}_{\circledcirc}) = E_{\phi_j}(\vec{\alpha}_{\circledcirc}) +F_{\phi_j}(\vec{\alpha}_{\circledcirc}) \ , 
\end{equation} 
i.e. the Hartree-Fock Hamiltonian at the target value consists of the zero-body and one-body terms of the target Hamiltonian normal-ordered with respect to the Hartree-Fock state at snapshot value $\vec{\alpha}_j$. 

We note that the Hamiltonian kernel in Eq.~(\ref{eq:hamiltonian}) is not symmetric, and the Fock matrix, $F_{\phi_j}(\vec{\alpha}_{\circledcirc})$, is not diagonal for $\vec{\alpha}_{\circledcirc} \neq \vec{\alpha}_{j}$. This is an important point and it ensures that---at a snapshot value---the solution of the generalized eigenvalue problem is indeed an eigenstate of the Hartree-Fock Hamiltonian (and not a GCM solution that is lower in energy).
The norm kernel for non-orthogonal reference states is given by~\cite{lowdin1955}
\begin{equation}
\langle \Phi_0 \vert \mathcal{O}_{ij} \vert \Phi_0\rangle = \mathop{\rm det}\left({\mathcal{O}_{ij}^{hh}}\right) \ ,
\end{equation}
here $\mathcal{O}_{ij}^{hh}$ is the matrix of overlaps between occupied (hole) states in $\langle \phi_i\vert $ and $\vert \phi_j\rangle$. To evaluate the Hamiltonian kernel we utilize the Thouless theorem~\cite{thouless1960} and write, 
\begin{equation}
\label{eq:thouless}
\langle \Phi_0 \vert \mathcal{O}_{ij} = \langle \Phi_0 \vert \mathcal{O}_{ij} \vert \Phi_0 \rangle \langle \Phi_0 \vert e^V \ ,
\end{equation}
with $V$ being a $1p$--$1h$ de-excitation operator. The matrix elements of $V$ in the hole-particle ($hp$) space is given by the matrix product~\cite{qiu2017}, 
\begin{equation}
V^{hp} = \left( \mathcal{O}_{ij}^{hh}\right)^{-1} \mathcal{O}_{ij}^{hp} \ .
\end{equation}
Inserting equation~(\ref{eq:thouless}) into equation~(\ref{eq:hamiltonian}) we obtain the algebraic equation, 
\begin{equation}
 \langle {\phi}_i \vert H_{\rm HF}(\vec{\alpha}_{\circledcirc}) \vert \phi_j \rangle  =  
 \langle {\phi}_i \vert \phi_j \rangle  \left( E_{\rm HF} + \sum_{hp} V^h_p F_h^p \right) \ .  
\end{equation}
Here $E_{\rm HF}$ and $F$ are the vacuum energy and one-body normal ordered terms, respectively, of $H_{\rm HF}(\vec{\alpha}_{\circledcirc})$ with respect to $\vert \phi_j \rangle$.

The norm and Hamiltonian Hartree-Fock kernels can also be evaluated using a generalized Wick's theorem~\cite{lowdin1955,hoyos2012,robledo2020,burton2021}. We verified that this alternative approach gives results that agree with the one used in this work. Having obtained the reduced-order model for the target Hartree-Fock state by diagonalizing the generalized non-symmetric eigenvalue problem in equation~(\ref{eq:general_eigenvalue}), we evaluate the projected target Hartree-Fock energies from
\begin{equation}
E^{(J)}_{\circledcirc} = \frac{\langle \phi_{\circledcirc} \vert P_J H(\vec{\alpha}_{\circledcirc}) \vert \phi_{\circledcirc} \rangle} {\langle \phi_{\circledcirc} \vert P_J \vert \phi_{\circledcirc}  \rangle } \ .
\end{equation}
Here the full target Hamiltonian in equation~(\ref{eq:full_hamiltonian}) enters, and $P_J$ is the projection operator.

\section{Results}
\label{sec:results}
Ab initio computations of realistic ground-state energies for spherical light- and medium-mass nuclei are demanding calculations but can nowadays be performed routinely~\cite{hergert2020}. In this work, however, we focus on our novel results for deformation and shape-coexistence emerging in a multiscale setting encompassing both small excitation energies and large total binding energies. 

\subsection{Multiscale physics of neutron-rich neon nuclei}
\label{sec:neons}


\begin{figure*}
    \includegraphics[width=1.0\textwidth]{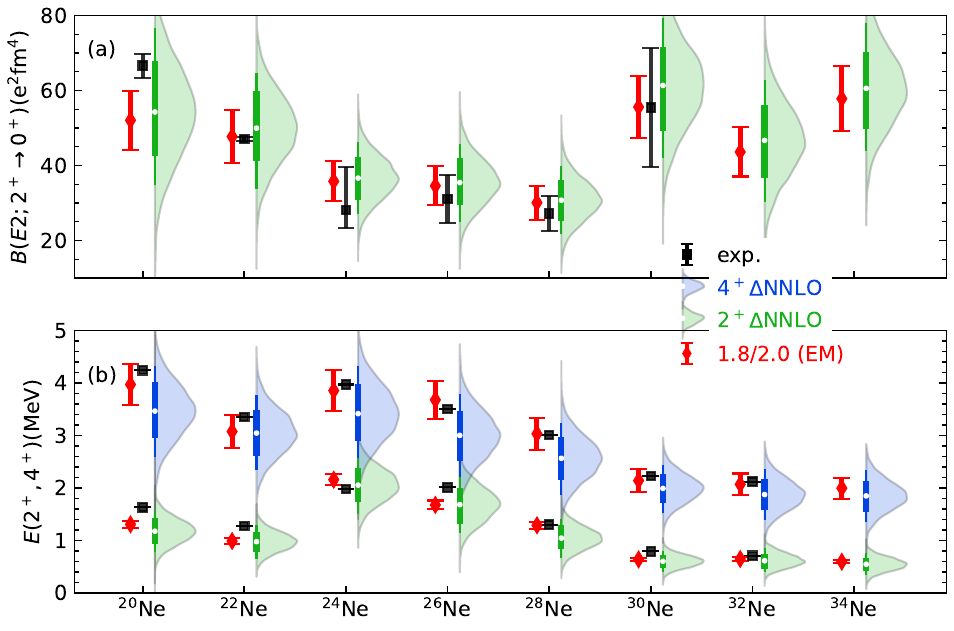}
    \caption{The upper panel (a) shows electric quadrupole transition strengths from the first excited $2^+$ state to the ground-state while the lower panel (b) shows energies of the lowest $2^+$ and $4^+$ states in the even nuclei $^{20-34}$Ne. Theoretical results are computed using angular-momentum projected coupled-cluster. Point predictions using the interaction 1.8/2.0(EM)~\cite{hebeler2011} (red diamonds with uncertainty estimates from many-body method and model-space truncations) are shown together with full posterior predictive distributions using the delta-full NNLO interaction ensemble ($\Delta$NNLO), including sampling of method and model errors. For the latter, 68\% and 90\% credible intervals are shown as a thick and thin vertical bar, respectively, and the median is marked as a white circle. Experimental data~\cite{ensdf} (and  Refs.~\cite{doornenbal2009,murray2019} for $^{32}$Ne) are shown as black squares with error bars.}
    \label{fig:neons}
\end{figure*}

Let us study first the collective properties of neon nuclei. The results for even-mass isotopes  are shown in Fig.~\ref{fig:neons}. The lower panel shows the  excitation energies $E(2^+)$ and $E(4^+)$ of the lowest spin-parity $J^\pi=2^+$ and $4^+$ states, respectively, in the $^{20-32}$Ne compared with available experimental data. The results based on the 1.8/2.0(EM) interaction are marked by red diamonds and include uncertainty estimates from method and model-space truncations. The relative accuracy of our calculations with the interaction 1.8/2.0(EM) is about 2--3\%. The distributions labelled $\Delta$NNLO are posterior predictive distributions given by the set
\begin{equation}
    \left\{ y_k(\vec\alpha) + \varepsilon_\mathrm{MB} +  \varepsilon_\mathrm{EFT} : \vec\alpha \sim p(\vec\alpha | \mathcal{D}_\mathrm{cal}) \right\},
\end{equation}
where $y_k$ is the ab initio model prediction for observable $k$, while $\varepsilon_\mathrm{MB}$ and $\varepsilon_\mathrm{EFT}$ are samples from the stochastic models for errors due to the many-body method and model-space truncation and the truncation of the effective field theory expansion, respectively (see App.~\ref{app:error} for details). When generating the posterior predictive distributions we used $10,000$ samples from the posterior $p(\vec\alpha | \mathcal{D}_\mathrm{cal})$ obtained via importance resampling (see Sec.~\ref{sec:hamiltonian}).

Theoretical results are consistent with each other and accurately reproduce the experimental trend where data exist. For neon nuclei with neutron number below $N=20$, i.e., $^{30}$Ne, the $\Delta$NNLO posterior predictive distributions indicate somewhat too compressed spectra. For neutron number $N\ge 20$, the excitation energies follow the pattern $E(J)= J(J+1)/(2\Theta)$ of a rigid rotor, and the relatively small values reflect a large moment of inertia $\Theta$ and a strong deformation. The computed energy-ratio $R_{42}\equiv E(4^+)/E(2^+)$ 
values of $^{34}$Ne are $3.37 \pm 0.13$ from the posterior (68\% credible interval) and $3.38 \pm 0.12$ for 1.8/2.0(EM); both are close to the value $10/3$ of a rigid rotor. Results for the dripline nucleus $^{34}$Ne are predictions.

The upper panel of Fig.~\ref{fig:neons} shows the computed electric quadrupole transition strength from the first excited $2^+$ state to the $0^+$ ground-state.  Overall, theory agrees with data, though both theoretical and experimental uncertainties are substantial. For $N=20$, theory is as precise as data, and we make equally precise predictions for $^{32,34}$Ne.

These results demonstrate that the inclusion of short- and long-range correlations on top of an axially deformed reference state enables to accurately capture quadrupole collectivity. The resulting picture is simple and well aligned with ideas from effective field theories where short-range and long-range correlations are distinguished. The symmetry-breaking Hartree-Fock state  provides us with a leading-order description of the nucleus and yields the Fermi momentum as a dividing scale. Short-range (high-momentum) contributions are included via few-particle--few-hole excitations within standard coupled-cluster theory; this yields almost all binding energy. Long-range correlations enter as a higher-order correction via many-particle--many-hole excitations and are included by symmetry projection of the correlated and symmetry-broken coupled-cluster state. This contributes little energy to the binding of the nucleus but is essential for its collective structure. This combined approach overcomes a long-standing multiscale challenge in low-energy nuclear physics~\cite{brown2006,parzuchowski2017,otsuka2020,miyagi2020,stroberg2022}. 

We also computed rotational bands for magnesium isotopes. These nuclei are much better known than the neon isotopes. For this reason, we limited ourselves to the the 1.8/2.0(EM) potential and only performed projected Hartree-Fock calculations. The result are close to data and shown for completeness in App.~\ref{app:mg}.  

\subsection{Shape coexistence in $^{30}$Ne and $^{32}$Mg}

The nuclei $^{30}$Ne and $^{32}$Mg are particularly interesting as they contain 20 neutrons, which is a magic number in the traditional shell model~\cite{mayer1955}. Though these nuclei are deformed in their ground-state~\cite{yanagisawa2003,doornenbal2016}, signatures of the $N=20$ magic number can be seen in our calculations and lead to shape coexistence. The following calculations are based on the 1.8/2.0(EM) interaction.

\begin{figure*}[htb]
    \includegraphics[width=0.99\textwidth]{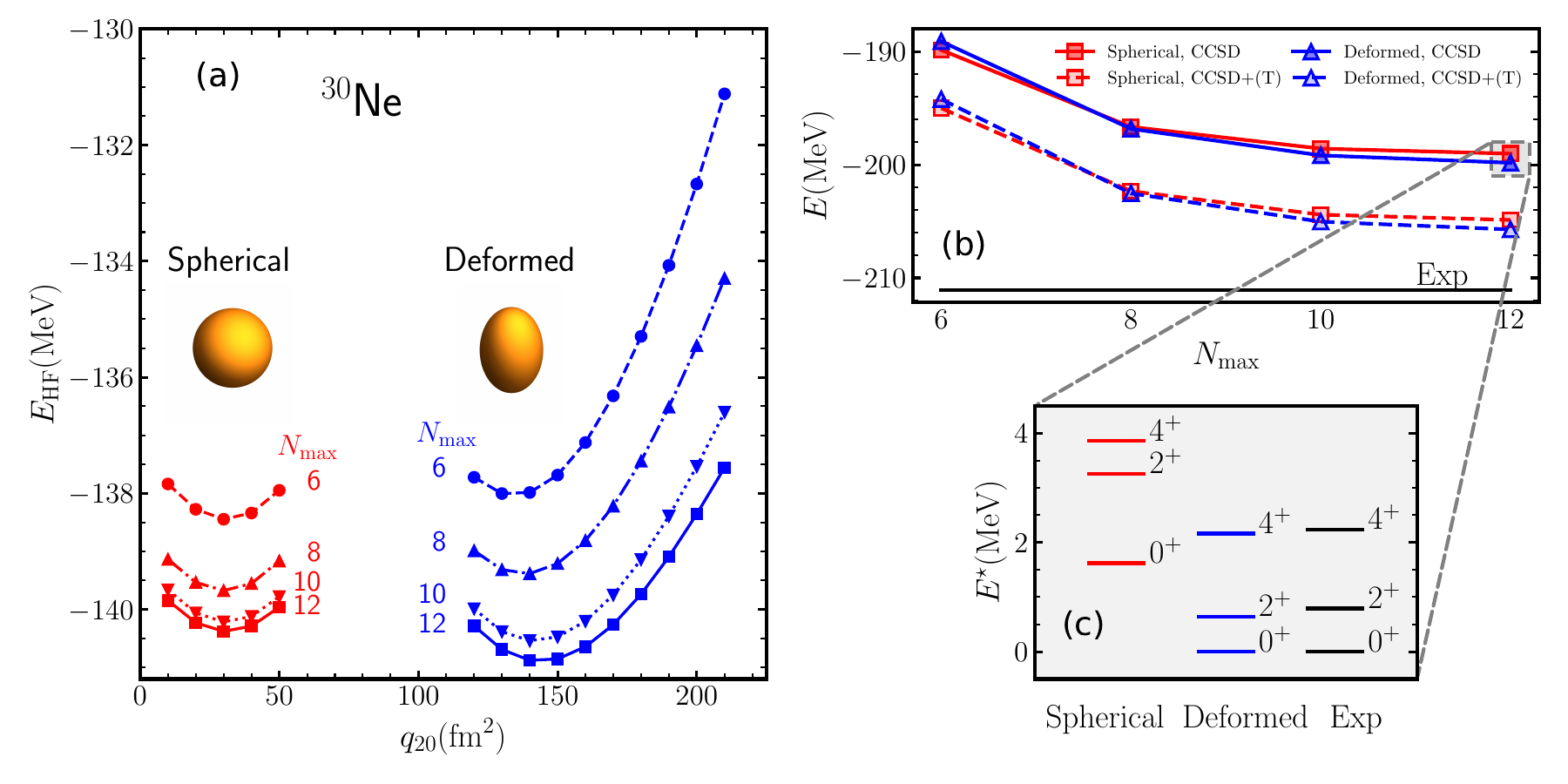}
    \caption{ 
    Panel (a) shows the Hartree-Fock energy for $^{30}\text{Ne}$ as a function of the quadrupole deformation $q_{20}$ for the nearly spherical (left, red curves) and deformed (right, blue curves) states for increasingly larger model spaces (labelled by $N_{\rm max}$) with an oscillator frequency of $\hbar \omega = 14$~MeV. Panel (b) demonstrates the convergence of the ground-state energy from symmetry-breaking coupled cluster with singles and doubles (CCSD) and triples estimates [CCSD+(T)]. These deformed states are superpositions of states with good angular momentum. Panel (c) shows the rotational bands obtained from angular momentum projection of the deformed states in a model space $N_{\rm max}=8$. All calculations are based the interaction 1.8/2.0(EM).}
    \label{fig:ne30_benchmark}
\end{figure*}

\begin{figure*}[htb]
\includegraphics[width=0.99\textwidth]{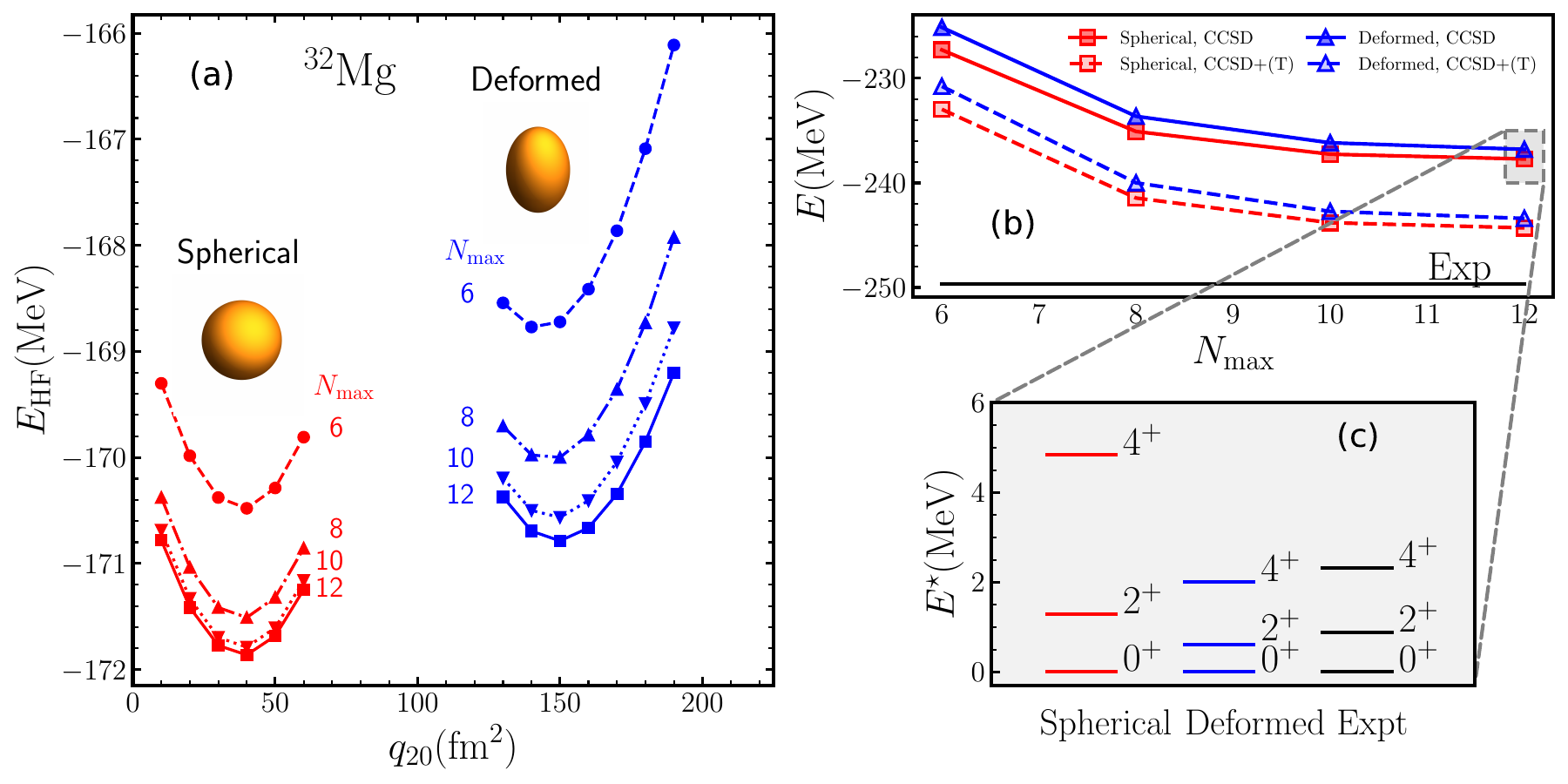}
   \caption{
   The same as Fig.~\ref{fig:ne30_benchmark} but for $^{32}\text{Mg}$.
    \label{fig:mg32_benchmark}}
\end{figure*}

For $^{30}$Ne and $^{32}$Mg we perform constrained quadrupole-moment Hartree-Fock calculations, as described in Sec.~\ref{sec:no2b}, starting from a spherical harmonic oscillator basis with oscillator frequency $\hbar \omega = 14$~MeV. In both nuclei we find one minimum corresponding to a nearly spherical shape plus a second one with strong deformation, see Figs.~\ref{fig:ne30_benchmark} and ~\ref{fig:mg32_benchmark}, Panel (a). 
The near-spherical configurations are close in energy to the well-deformed ones. This suggests that the shape coexistence, observed for $N=20$ in $^{32}$Mg~\cite{wimmer2010}, remains present when two protons are removed from that nucleus. We note that the nearly spherical and well deformed Hartree-Fock states exhibit different occupations of Nilsson orbitals. The nearly spherical states reflect the $N=20$ sub-shell closure for neutrons. For the deformed states two intruder orbitals with $j_z = \pm 1/2$ from the $pf$-shell are occupied. As we work in a single-reference framework, there is no continuous connection between the two different Hartree-Fock energy ``surfaces."

Panels (b) and (c) of Figs.~\ref{fig:ne30_benchmark} and ~\ref{fig:mg32_benchmark} demonstrate the vast difference of scales of the computed binding energies $E$ and the rotational excitation energies $E^*$. Both scales simultaneously emerge in our calculations. The main contribution to the ground-state energy comes from short-range correlations in the wave function. Two-particle--two-hole excitations, included via CCSD, give the main contribution to the correlation energy [i.e. the energy in excess of the Hartree-Fock energy shown in Panel~(a)]. Here, the triples estimates are taken as 10\% of the CCSD correlation energy, which was confirmed in benchmark calculations, see App.~\ref{app:benchmark}.
We see that the nearly spherical $0^+$ ground-state in $^{30}$Ne resides about 1.8~MeV above the prolate $0^+$ ground-state. 
Binding energies are reproduced within about 3\%, and angular momentum projection would further reduce the small discrepancy. 
In contrast to $^{30}$Ne, the energy difference between the competing spherical and prolate minima in $^{32}$Mg is only few tens of keV after angular-momentum projection. Thus, from our computations we can not conclusively decide which state corresponds to the ground-state in this nucleus.

\subsection{Global sensitivity analysis of deformation}
\label{sec:gsa}
The ab initio computations presented in this work have a high resolution that allows to study how individual terms corresponding to the 17 low-energy constants of a delta-full chiral interaction at NNLO~\cite{jiang2020}, with regulator cutoff 394 MeV$/c$, impact nuclear deformation. To quantify this, we perform a variance-based global sensitivity-analysis~\cite{sobol2001} of the $E(4^+)/E(2^+)$ ratio $R_{42}$ in $^{20,32}$Ne and $^{34}$Mg. It is sufficiently accurate to solve for the excited-state energies $E(2^+)$ and $E(4^+)$ in $^{20,32}$Ne and $^{34}$Mg using projection-after-variation Hartree-Fock (see Refs.~\cite{Frosini:2021sxj,hagen2022} and App.~\ref{app:benchmark}). However, the Monte Carlo sampling in a global sensitivity analysis requires prohibitively many projected Hartree-Fock computations. Indeed, we find it necessary to use one million samples to keep sampling uncertainties under control. To overcome this computational barrier we developed fast and accurate emulators for the excitation energies using the method described in Sec.~\ref{sec:rom}.

\begin{figure*}[!htbp]
    \includegraphics[width=0.32\linewidth]{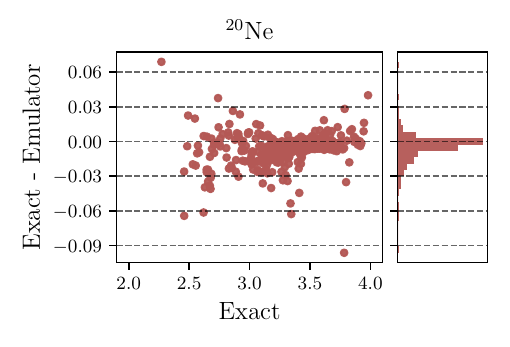}
    \includegraphics[width=0.32\linewidth]{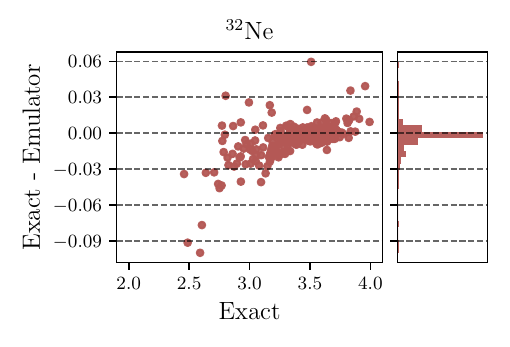}
    \includegraphics[width=0.32\linewidth]{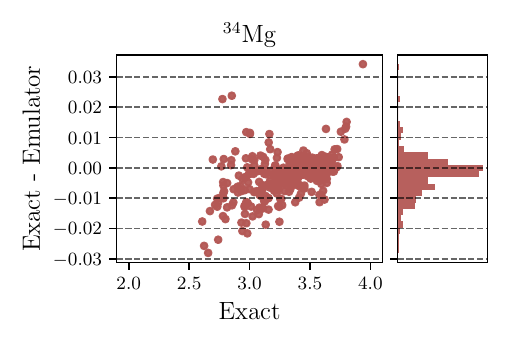}
    \caption{Accuracy of $R_{42}$ emulators as measured by the difference with respect to 400 exact Hartree-Fock calculations, on the $y-$axis, as a function of $R_{42}$ on the $x-$axis. Shown are the subsets of results for which $R_{42}\in[2,4]$, i.e., the range relevant to the global sensitivity analysis. The horizontal dashed lines indicate precision as they are spaced by two standard deviations of the projected histograms (shown in side panels).}
    \label{fig:xval_gsa}
\end{figure*}

The strategy for training the emulator is similar to Ref.~\cite{ekstrom2019}. We generated 68 snapshots of the first excited $2^+$ and $4^+$ states, in the Hartree-Fock approximation, using values for the 17 low-energy constants according to a space-filling latin hypercube design encompassing 20-30\% variation of their $\Delta$NNLO$_\text{GO}(394)$ values~\cite{jiang2020}. Figure~\ref{fig:xval_gsa} shows the accuracy of the resulting emulator for $R_{42}$ as quantified by comparison with 400 exact projected Hartree-Fock calculations. The standard deviation of the differences indicates a relative precision of $1\%$ (at the one sigma level) for the interval of $R_{42}$ values relevant to the global sensitivity analysis presented below. The excitation energies using our reduced-order model are accurate on the 10~keV level.

In the global sensitivity analysis we numerically quantify, and decompose, the variances of the excitation energies for the$^{20,32}$Ne and $^{34}$Mg isotopes due to sampling one million different values of the low-energy constants at NNLO. The variances are decomposed in terms of the leading $S$-wave contacts $\tilde{C}_{^3S_1}$ and $\tilde{C}_{^1S_0}^{(\tau)}$ with $\tau=nn, np, pp$ denoting the isospin projections, the subleading contacts $C_{^1S_0}$, $C_{^3S_1}$, $C_{^3P_0}$, $C_{^1P_1}$, $C_{^3P_1}$, and $C_{^3P_2}$ (acting in a partial wave as indicated by the subscript), and $C_{E_1}$ acting in the the off-diagonal triplet $S-D$ channel. We also include the four subleading pion-nucleon couplings $c_{1,2,3,4}$, as well as the $c_D$ and $c_E$ couplings governing the strengths of the short-range three-nucleon potential. The variance integrals underlying the sensitivity analysis are evaluated on a hypercubic domain centered on the $\Delta$NNLO$_\text{GO}(394)$ parameterization~\cite{jiang2020}. The size of the domain is based on recent Bayesian analyses~\cite{Wesolowski:2021cni,Svensson:2023twt} and naturalness arguments from effective field theory. In detail, we use $\pm 0.05$~GeV$^{-1}$ as the relevant range for each of the pion-nucleon couplings $c_i$ and $\pm 0.05\times 10^2$~GeV$^{-2}$ for the sub-leading constants $C_i$. The leading-order contact couplings $\tilde{C}_i$ are somewhat small, and their intervals are limited to $\pm 0.005\times 10^4$~GeV$^{-4}$. We examine our results for robustness by re-scaling all side-lengths of the hypercube by factors of $1/2$ and $2$. Even larger domains result in noticeable higher-order sensitivities which we did not analyze further. 

\begin{figure}[t]
    \includegraphics[width=0.99\linewidth]{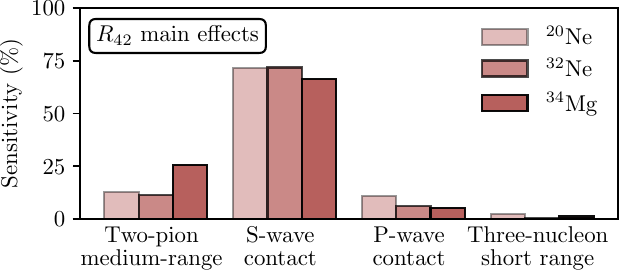}
    \caption{ 
    Global sensitivity-analysis of $10^6$ emulations of projected Hartree-Fock computations in $^{20,32}$Ne and $^{34}$Mg using delta-full chiral effective field theory at next-to-next-to leading order. 
    \label{fig:gsa}}
\end{figure}

A majority of the samples in all three nuclei have $R_{42} \approx 10/3$ within 5\%. This indicates that an axially-deformed rigid rotor and emergent symmetry-breaking is a robust feature of the effective field theory description of these nuclei. The variance of the conditional mean of $R_{42}$, with respect to a low-energy constant, divided by the total variance is a dimensionless ratio called the \textit{main effect}. Overall, we find that more than $90\%$ of the variance in $R_{42}$ is explained via main effects. Figure~\ref{fig:gsa} shows the main effects for $R_{42}$ in $^{20,32}$Ne and $^{34}$Mg in terms of groups of low-energy constants proportional to medium-range two-pion exchange, short-range nucleon-nucleon contact interactions in the $S$- and $P$-waves, and the short-range three-nucleon interactions consisting of a contact-interaction and pion-exchange plus contact interaction. A greater value of the main effect indicates a larger sensitivity of $R_{42}$ to the corresponding component of the chiral interaction. For all three nuclei, more than $50\%$ of the deformation is driven by the $S$-wave contact part of the interaction. Adding short-range repulsion appears to increase deformation, probably by reducing pairing. Medium-range two-pion exchange is also important. Increasing its strength increases deformation, presumably by adding attraction in higher partial waves.

\begin{figure*}[htb]
    \includegraphics[width=0.99\textwidth]{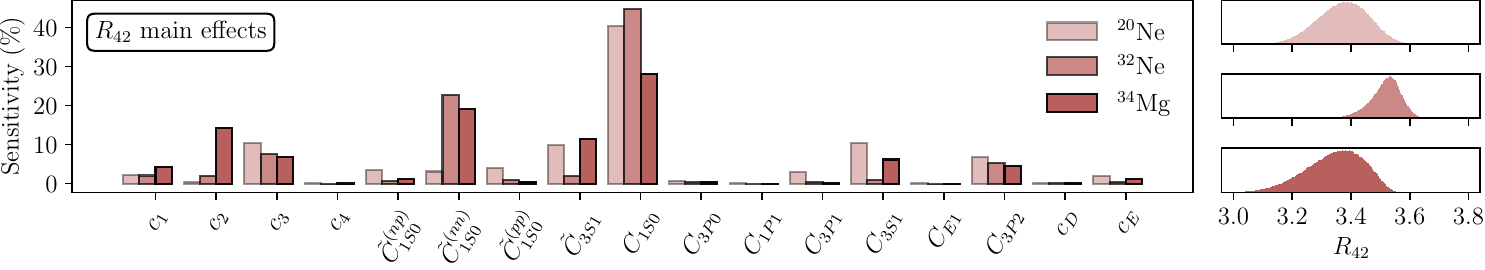}
    \caption{
    Main effects for the $R_{42}$ deformation measure in $^{20,32}$Ne and $^{34}$Mg from the low-energy constants as obtained in a global sensitivity-analysis of $10^6$ emulations of projected Hartree-Fock computations using delta-full chiral effective field theory at next-to-next-to leading order. Distributions of $R_{42}$ for $^{20}$Ne, $^{32}$Ne, and $^{34}$Mg are also shown.}
    \label{fig:gsa_R42_detail}
\end{figure*}

Figure~\ref{fig:gsa_R42_detail} shows the main effects for $R_{42}$ in $^{20,32}$Ne and $^{34}$Mg in more detail. For all three nuclei, about $40\%$ of the deformation is driven by the subleading pion-nucleon coupling $c_3$ and the subleading singlet $S$-wave contact $C_{^1S_0}$. The $c_3$ coupling enters the attractive central part from the medium-range two-pion exchange in the nucleon-nucleon potential, and also in the three-nucleon potential~\cite{krebs2007}. For $^{32}$Ne and $^{34}$Mg, with many more neutrons than $^{20}$Ne, deformation becomes more sensitive to the isospin-breaking $S$-wave contact in the neutron-neutron channel. For $^{34}$Mg, the ratio $R_{42}$ appears to become more sensitive to the the pion-nucleon coupling $c_2$. See Appendix~\ref{app:gsa} for the sensitivities of Hartree-Fock energies, including ground-state energies.

We can also use the posterior samples employed in  Fig.~\ref{fig:neons} to probe what impacts deformation. The most relevant parts of the nuclear interaction can then be identified by studying correlations between the observable $R_{42}$ in $^{32}$Ne and individual low-energy constants. We find that the correlation is strongest for the $S$-wave contact term (with a correlation coefficient $r=0.73$, i.e., an increase in the repulsive low-energy constant increases deformation), but it is  also sizeable for the three-nucleon contact interaction (see Appendix~\ref{app:gsa} for details). Comparing these results with the conditional variances from the global sensitivity analysis confirms the importance of pairing via the $^1S_0$ channel. We note that the domain of low-energy constants used in the global sensitivity analysis is smaller than the region spanned by the Bayesian posterior interaction ensemble. 

Our sensitivity analysis is a first step and the emerging picture is still incomplete. Indeed, we did not gauge the sensitivities of tensor and spin-orbit terms in the Hamiltonian that are independent of the low-energy constants. We also remind the reader that our potential lacks $D$-wave contacts beyond the off-diagonal triplet $S-D$ coupling, because these enter only at the next higher order. It will be interesting to compare more complete results with shell-model pictures about what drives nuclear deformation~\cite{federman1979,nazarewicz1994,duflo1999,otsuka2006,poves2016}.

\section{Summary and Discussion}
\label{sec:summary}
In summary, we demonstrated how ab initio computations of nuclei can accurately describe binding energies, rotational bands, and collective electromagnetic transition strengths. These results were obtained in a non-perturbative  framework where dynamical correlations were included via coupled-cluster theory and static correlations via angular-momentum projection. These advances allowed us to explore how collective nuclear properties are sensitive to specific terms in effective Hamiltonians of low-energy quantum chromodynamics that include nucleon-nucleon and three-nucleon forces. We found that the contacts in the $^1S_0$ partial wave and the three-body contact play a major role in shaping nuclei. Using an ensemble of calibrated interactions, we made predictions with quantified uncertainties in neutron-rich neon nuclei. The uncertainties---while still considerable---are on par with experiment for the neon isotopes close to the neutron dripline. In particular, we predict shape coexistence in $^{30}$Ne.  

This work points to a conceptually simple and attractive multiscale picture of nuclei where the symmetry-breaking reference state contains the relevant physics aspects. Size-extensive methods then yield the lion share of the binding energy while symmetry-projection methods account for important collective components in the wave function that, however, contribute comparatively little to the binding energy. 

Our computations demonstrate the predictive power of ab initio methods. This reductionist approach combines ever-increasing computational modeling capabilities and heuristic techniques to capture dynamical and emergent properties of complex systems. Other collective degrees of freedom (besides rotations) such as vibrations or competing shapes can be added in the same framework because they can be realized as exponentiated one-body operators. Our sensitivity analysis provided first insights to the link between microscopic nuclear forces and complex nuclear phenomenology. The ab initio methods, and accompanying emulator techniques, developed in this work open for computational statistics analyses to identify principal components that drive emergent phenomena in finite systems. While we focused on an important problem in nuclear physics, similar challenges exist in quantum chemistry. Thus, we expect the novel symmetry-projection techniques of correlated states to be useful in many other applications.

\begin{acknowledgments}
This work was supported (in part) by the U.S. Department of Energy, Office of Science, Office of Advanced Scientific Computing Research and Office of Nuclear Physics, Scientific Discovery through Advanced Computing (SciDAC) program (SciDAC-5 NUCLEI);  by the U.S. Department of Energy, Office of
Science, Office of Nuclear Physics, under Award Nos.~DE-FG02-96ER40963, and by the Quantum Science Center, a National Quantum Information Science Research Center of the U.S. Department of Energy; by the European Research Council (ERC) under the European Unions Horizon 2020 research and innovation program (Grant Agreement No. 758027); by the Swedish Research Council (Grants No.~2017-04234, No.~2020-05127 and No.~2021-04507). Computer time was provided by the Innovative and Novel Computational Impact on Theory and Experiment (INCITE) programme. This research used resources of the Oak Ridge Leadership Computing Facility located at Oak Ridge National Laboratory, which is supported by the Office of Science of the Department of Energy under contract No. DE-AC05-00OR22725 and resources provided by the Swedish National Infrastructure for Computing (SNIC) at Chalmers Centre for Computational Science and Engineering (C3SE), and the National Supercomputer Centre (NSC) partially funded by the Swedish Research Council through grant agreement No.~2018-05973.    
\end{acknowledgments}

\clearpage

\appendix
\section{Overview}
The appendices contain a number of details that support and further illuminate the results presented in the main text. Appendix~\ref{app:benchmark} presents benchmarks and details about model-space dependencies for neon nuclei. Appendix~\ref{app:mg} presents details regarding results in magnesium isotopes. We state our assumptions about various uncertainties in App.~\ref{app:error}. Finally, we give many more details about our global sensitivity analysis in App.~\ref{app:gsa}.

\section{Benchmarks, model-space dependence, and ground-state energies for neon isotopes}
\label{app:benchmark}

For the computation of the ground-state energies of the $^{20-34}$Ne isotopes we use the 1.8/2.0(EM) interaction and follow the approach in~\cite{novario2020} and use a natural orbital basis and the coupled-cluster with singles-doubles and leading-order triples excitations, known as the CCSDT-1 approximation~\cite{lee1984,watts1995}. The use of natural orbitals allows for converged CCSDT-1 calculations by imposing a cut on the product of occupation numbers for three particles above the Fermi surface and for three holes below the Fermi surface~\cite{novario2020}. We used a model-space of 13 major oscillator shells with the oscillator frequency $\hbar \omega = 14$~MeV. 

\begin{figure}[b]
\includegraphics[width=0.99\linewidth]{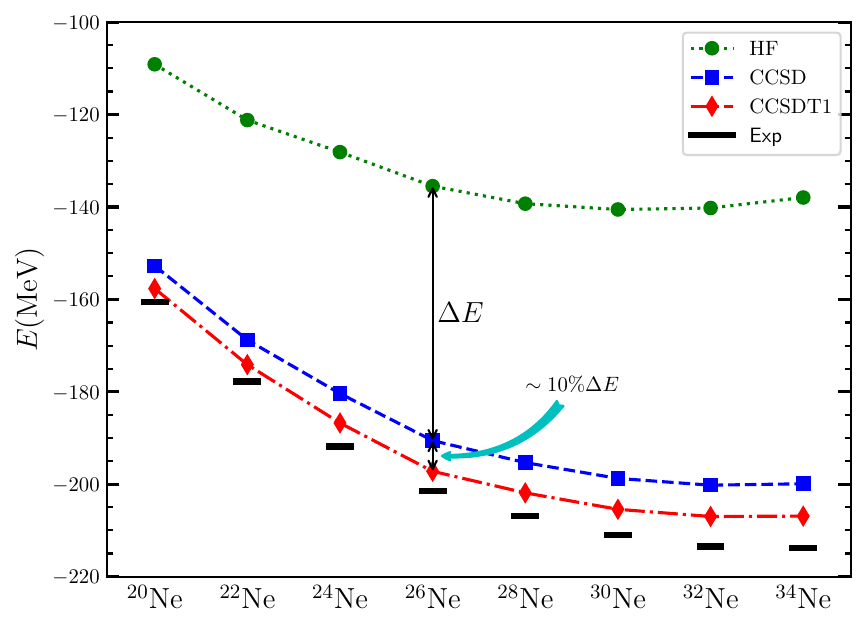}
    \caption{ 
    Energies from the Hartree-Fock, CCSD and CCSDT-1 approximations using the interaction 1.8/2.0(EM) from chiral effective field theory~\cite{hebeler2011}. Results are based on a natural orbital basis built from 13 major oscillator shells with a frequency of $\hbar\omega = 14$~MeV. The triples correction amounts to about 10\% of the correlation energy from CCSD ($\Delta E$).}
    \label{fig:neons-be-magic}
\end{figure}

Figure~\ref{fig:neons-be-magic} shows that binding energies are reproduced within about 3\%. Angular momentum projection is expected to further reduce the small discrepancy. We note that the triples correlation energy for all neon isotopes amounts to about 10\% of the correlation energy from CCSD. This is consistent with findings for coupled-cluster computations of closed-shell spherical nuclei~\cite{hagen2014} and in quantum chemistry~\cite{bartlett2007}.   
This justifies the triples estimates presented in Figs.~\ref{fig:ne30_benchmark} and \ref{fig:mg32_benchmark} of the main text.

\begin{figure}[h]  \includegraphics[width=0.99\linewidth]{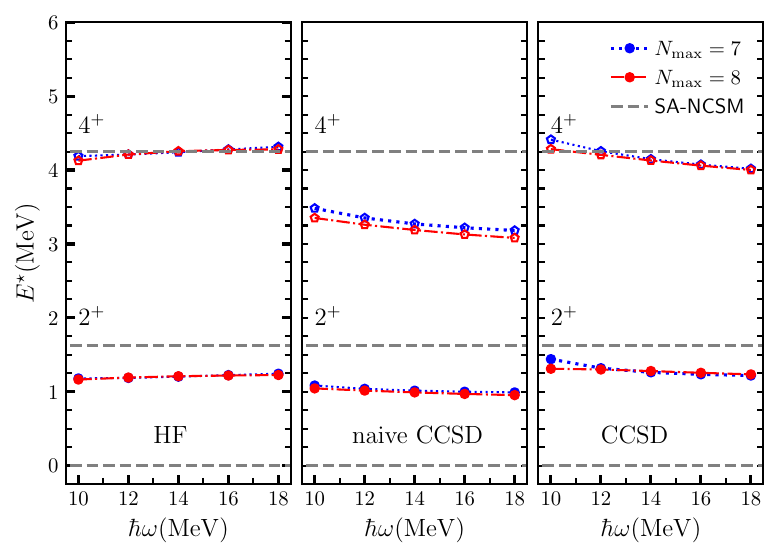}
    \caption{
    Comparison between projection-after-variation Hartree Fock (HF), naive projected coupled-cluster (naive CCSD), and projected coupled-cluster method (CCSD) for the excited $2^+$ and $4^+$ in $^{20}$Ne using the NNLO$_{\mathrm{opt}}$ nucleon-nucleon interaction. The dashed lines show benchmark results from the symmetry-adapted no-core shell-model (SA-NCSM)~\cite{launey2020}.}
    \label{fig:ne20_benchmark}
\end{figure}

We turn to benchmarks with the symmetry-adapted no-core shell model (SA-NSCM)~\cite{dytrych2020,launey2020,heller2022new}. These are based on the NNLO$_{\mathrm{opt}}$ nucleon-nucleon interaction~\cite{ekstrom2013} and the $^{20}$Ne nucleus.
Figure~\ref{fig:ne20_benchmark} shows the $2^+$ and $4^+$ rotational states as a function of the oscillator frequency $\hbar\omega$ for model spaces consisting of $N_{\rm max}+1$ shells. We compare  projection-after-variation results from Hartree-Fock with those from a ``naive'' (i.e. the left state is simply the adjoint of the reference state) and the bi-variational coupled-cluster ansatz in equation~\ref{eq:CC_PAV}, respectively. We see that both excited states are well converged with respect to the model-space size for Hartree-Fock. The coupled-cluster results exhibit a bit more model-space dependence.  This might be because the disentangled coupled-cluster approach does not restore the broken symmetry exactly~\cite{qiu2017,hagen2022}. We also see that projected  Hartree-Fock and the projected bi-variational coupled-cluster are close to each other and to the SA-NCSM results. The ``naive'' coupled-cluster approach yields more compressed spectra.  The agreement between Hartree-Fock and the much more expensive SA-NCSM and projected coupled-cluster results show how simple the physics behind rotational bands can be. This justifies the usage of a Hartree-Fock-based reduced-order model in the global sensitivity analysis of the ratio $R_{42}$. The accuracy of Hartree-Fock rotational bands is presumably limited to light nuclei where superfluidity and pairing correlations are less important.

\begin{figure}[H]
\includegraphics[width=0.99\linewidth]{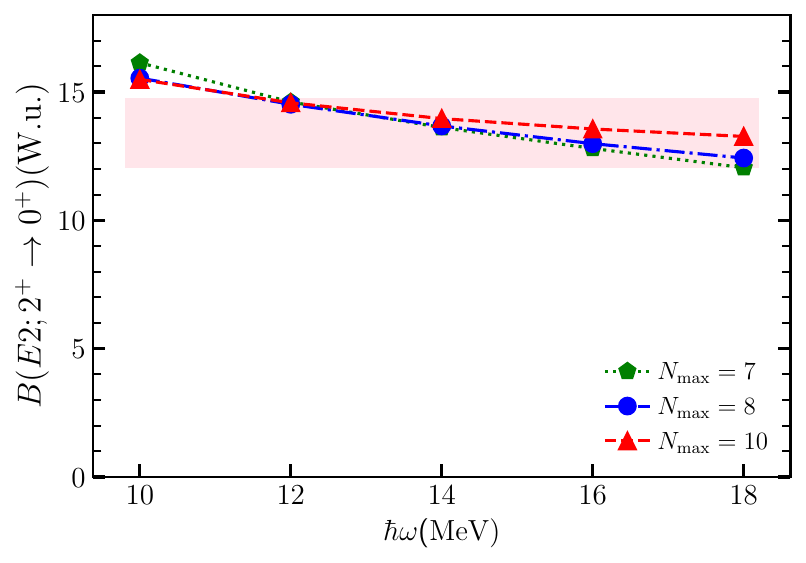}
    \caption{
    Results from projected coupled-cluster method (CCSD) using the NNLO$_{\mathrm{opt}}$ nucleon-nucleon interaction~\cite{ekstrom2013} for different model-space sizes ($N_{\rm max}$) and oscillator frequencies ($\hbar \omega$), and compared with the symmetry-adapted no-core shell-model (red band)~\cite{launey2020}.}
    \label{fig:ne20_be2_benchmark}
\end{figure}

We also want to benchmark the electric quadrupole transition strength. Our calculations are based on projected coupled-cluster theory, as described in Sec.~\ref{sec:em}. 
Figure~\ref{fig:ne20_be2_benchmark} compares our computation of the $B(E2)$ strength in $^{20}$Ne with the SA-NCSM results from Ref.~\cite{launey2020}, again for the  nucleon-nucleon interaction NNLO$_{\mathrm{opt}}$. Both results agree within uncertainties from finite model spaces. The benchmarks with the SA-NCSM give us confidence in the accuracy of our computations.

\begin{figure*}[htb]
\includegraphics[width=0.95\textwidth]{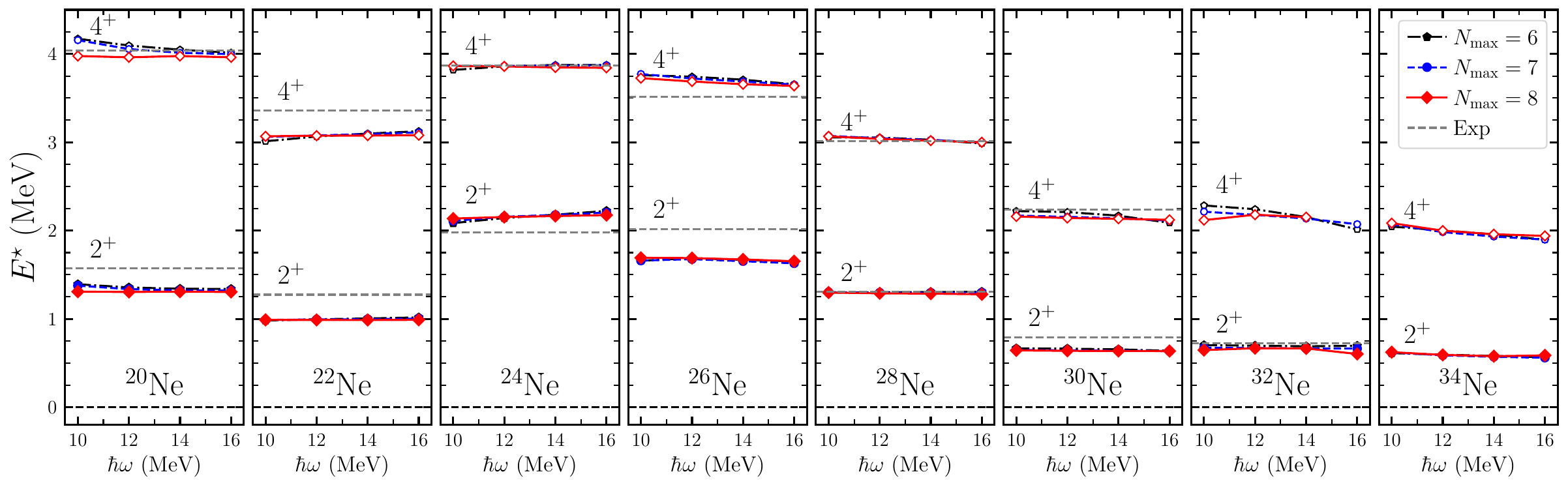}
    \caption{
    Energies $E$ of the lowest $2^+$ and $4^+$ states in even nuclei $^{20-34}$Ne, computed with the interaction 1.8/2.0(EM) from chiral effective field theory~\cite{hebeler2011}, shown as a function of the oscillator frequency and for various model spaces, and compared to data.}
    \label{fig:neons-magic}
\end{figure*}

We turn to details regarding the results shown in Fig.~\ref{fig:neons} of the main text and focus on the the interaction 1.8/2.0(EM).
Figure~\ref{fig:neons-magic} shows the energies of the $2^+$ and $4^+$ excited states in $^{20-34}$Ne  and compares them to data for different model-spaces (parameterized by the oscillator frequency $\hbar\omega$ and the number of shells $N_{\rm max+1}$). We find that the states are well converged with respect to model-space size. The variation of the results with respect to the model-space is shown as an uncertainty in Fig.~\ref{fig:neons} of the main text. Our results for $^{20}$Ne agree with those by \textcite{frosini2022} using the same interaction.

\begin{figure*}[htb]
    \includegraphics[width=0.95\textwidth]{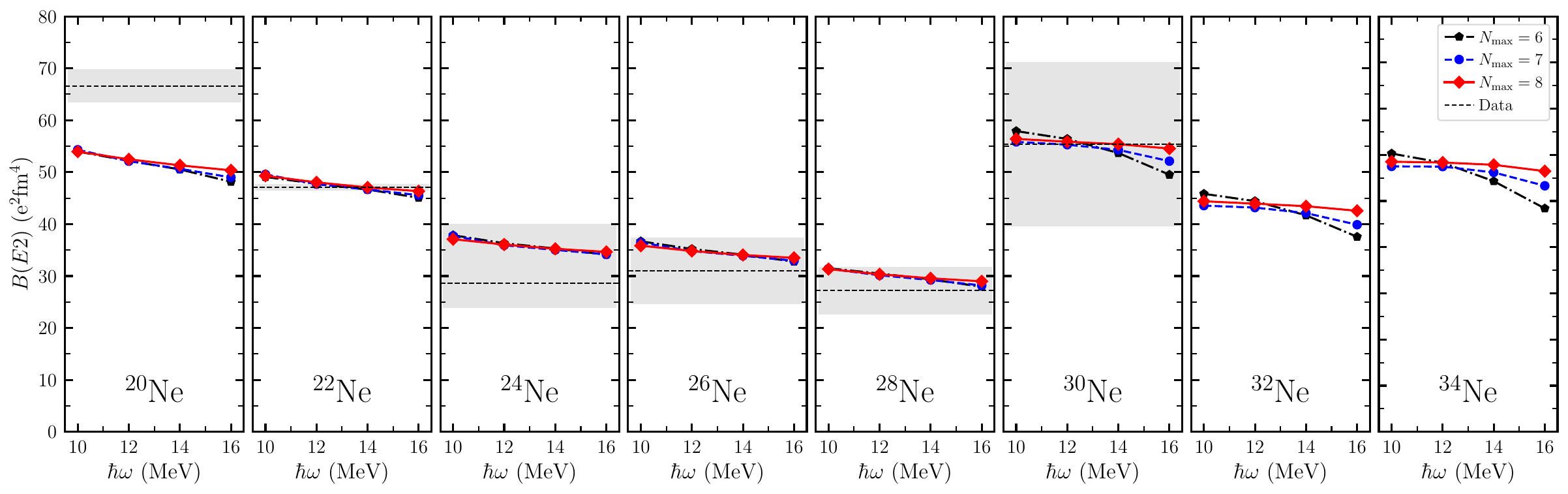}
    \caption{
    $B(E2, \downarrow)$ computed with the interaction 1.8/2.0(EM) from chiral effective field theory~\cite{hebeler2011} as a function of the oscillator frequency ($\hbar \omega$) for various model spaces sizes ($N_{\rm max}$). Results are compared with available data (black dotted lines with grey uncertainty bands), taken from Ref.~\cite{pritychenko2016} for $^{20,24-28}$Ne, from Ref.~\cite{HENDERSON2018468} for $^{22}$Ne, and from Ref.~\cite{doornenbal2016} for $^{30}$Ne. }
    \label{fig:neons-magic-be2}
\end{figure*}

Figure~\ref{fig:neons-magic-be2} shows the $B(E2,\downarrow)$ for the transition between the first excited $2^+$ state and the ground-state in $^{20-34}$Ne obtained with the interaction 1.8/2.0(EM) and compared to available data for different model-spaces (parameterized by the oscillator frequency $\hbar\omega$ and the number of shells $N_{\rm max+1}$). We find that the $B(E2,\downarrow)$ is well converged with respect to model-space size for isotopes with $N \ge 20$, while less so for neon nuclei with $N<20$. The variation of the results with respect to the model-space is shown as an uncertainty in the upper panel of Fig.~\ref{fig:neons} in the main text.

\section{{Uncertainty estimates}}
\label{app:error}
The results based on the 1.8/2.0(EM) interaction include estimates of the method uncertainty coming from the use of method and model-space truncations. To estimate the latter we considered the spread of results obtained for $N_{\rm max} = 6-8$ and $\hbar\omega = 10-16$~MeV (see Figs~\ref{fig:neons-magic-be2} and ~\ref{fig:neons-magic}). We find that model-space truncation errors are smaller than the estimated method truncation errors (see below). For the delta-full NNLO interaction model we employ the ensemble of Bayesian posterior samples and quantify both method and model uncertainties for predicted observables. We employ a fixed model-space of $N_{\rm max} = 7$ and $\hbar\omega = 14$~MeV and assign normally distributed method errors with relative (one sigma) errors of 10\% (5\%) for $2^+$ ($4^+$) excitation energies (corresponding to about $100-150$~keV). For the $B(E2,\downarrow)$ we assign a 15\% (one sigma) method error from our benchmark with the symmetry adapted no-core shell-model~\cite{dytrych2020} in $^{20}$Ne,  see Fig.~\ref{fig:ne20_be2_benchmark}. Moreover, we assign 10\% (one sigma) relative EFT truncation errors for all excitation energies and all transition strengths. We use experimental values to translate relative errors to absolute ones, and we use reference values of $1.0$ and $2.5$~MeV for $2^+$ and $4^+$ excitation energies, respectively, and $50$~$e^2\mathrm{fm}^4$ for $B(E2,\downarrow)$ to get absolute errors for $^{32,34}$Ne where experimental data is not available. All errors are described by independent  normal distributions.

\section{Details to  magnesium nuclei}
\label{app:mg}

Our studies of heavier magnesium nuclei are limited to using projection after variation of  Hartree-Fock states. 
This simplification is justified based on Ref.~\cite{hagen2022} and the comparison of the rotational bands obtained from Hartree-Fock and coupled-cluster theory in the $^{20}$Ne, as shown in Appendix~\ref{app:benchmark}.

Figure~\ref{fig:magnesiums} shows projected Hartree-Fock results for the energies $E(2^+)$ and $E(4^+)$ in $^{32-40}$Mg computed with the 1.8/2.0(EM) interaction as a function of the oscillator frequency $\hbar\omega$ in  model spaces consisting of $N_{\rm max}+1$ shells. The results exhibit only a small model-space dependence and are close to data. 

For  $^{32}$Mg the Hartree-Fock results already confirm the shape coexistence in this nucleus~\cite{wimmer2010}. While the spherical Hartree-Fock state is about 1~MeV lower in energy than the deformed one, the inclusion of short-range correlations via coupled-cluster theory reduces this difference. 

Similarly, we see shape coexistence in $^{40}$Mg, confirming the experimental \cite{crawford2019} and theoretical results~\cite{tsunoda2020}.
Our calculations of the dripline nucleus $^{40}$Mg include couplings to the particle continuum via a Woods-Saxon basis consisting of bound and scattering states for the neutron $p_{3/2}$ partial wave, following Ref.~\cite{hagen2016}. 

One expects an inversion of the $p_{3/2}$ and $f_{7/2}$ single-particle orbitals close to the magnesium dripline. This is supported by the observation that $^{37}$Mg is a deformed $p-$wave halo nucleus~\cite{kobayashi2014} and  mean-field computations accounting for deformation and continuum coupling~\cite{hamamoto2012,hamamoto2016}. Indeed our calculations for $^{38,40}$Mg show an inversion of the $K^\pi = 7/2^-$ and $K^\pi = 1/2^-$ single-particle orbitals (where $K$ denotes the single-particle angular-momentum component along the axial symmetry axis). We find that $^{34-40}$Mg are all prolate in their ground-state, and the computed rotational bands are close to data. 

Interestingly, for $^{40}$Mg we also find an oblate Hartree-Fock state that is close in energy to the prolate ground state. Performing coupled-cluster calculations for these two references we find that the oblate band head is about  3~MeV above the prolate ground-state, indicating an onset of shape coexistence and a possible interpretation of the third observed excited state~\cite{crawford2019}. This picture is also consistent with the Monte-Carlo shell-model computations of \textcite{tsunoda2020}. Figure~\ref{fig:magnesiums} shows both the prolate and oblate $2^+$ and $4^+$ states, and we observe that the rotational structure of these two bands are very similar and close to that of a rigid rotor.

\begin{figure*}[htb]
\includegraphics[width=0.9\textwidth]{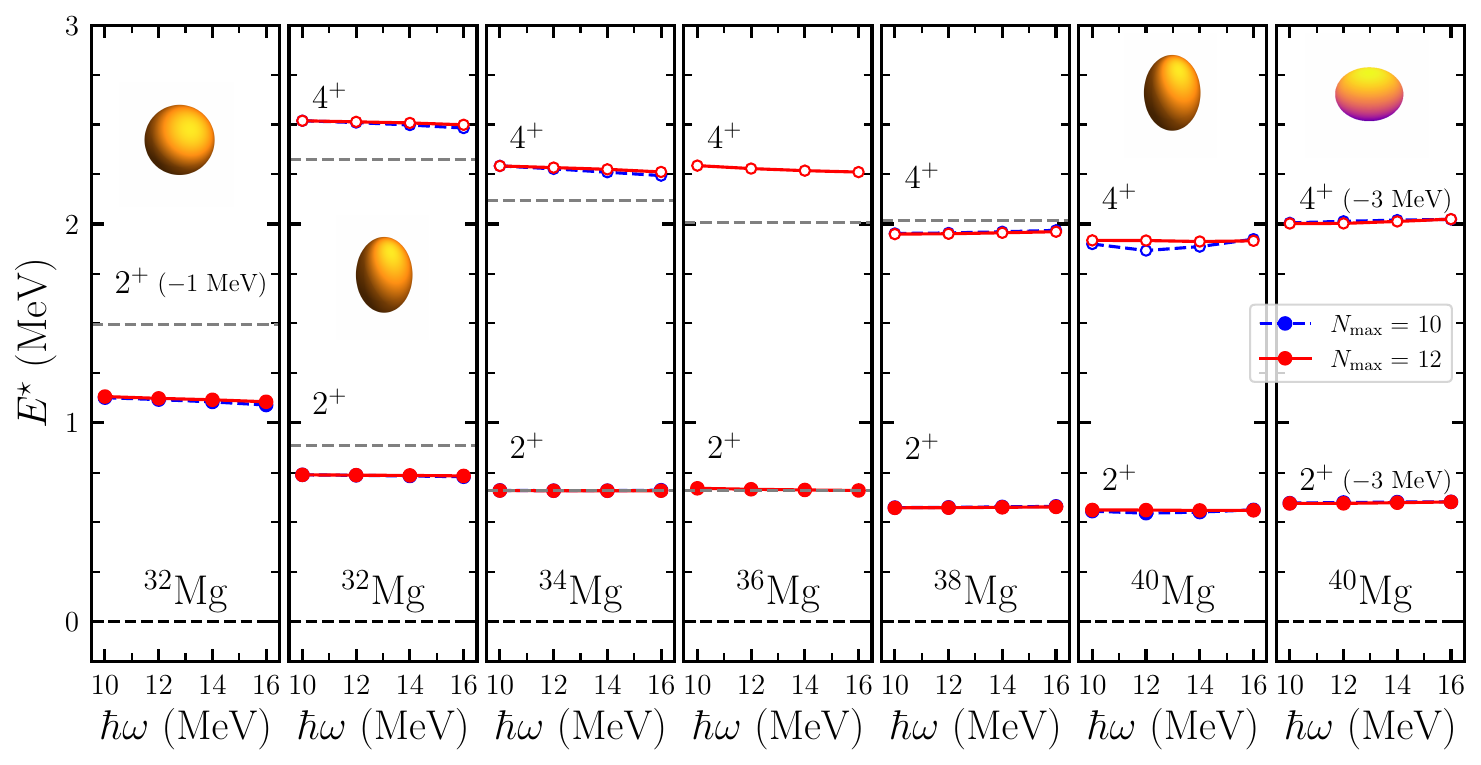}
    \caption{ 
    Energies as a function of the oscillator frequency ($\hbar \omega$) and for various model spaces ($N_{\rm max}$), computed using projected Hartree-Fock with the chiral interaction 1.8/2.0(EM)~\cite{hebeler2011} and compared to data (dashed horizontal lines). For $^{40}$Mg we show both the prolate and oblate rotational bands; the latter band head is about 3~MeV above the prolate ground state. The pictures show spherical, prolate deformed, and oblate deformed ellipsoids.}
    \label{fig:magnesiums}
\end{figure*}

\section{Sensitivity study for neon nuclei}
\label{app:gsa}

\begin{figure*}[!htb]
    \includegraphics[width=0.9\linewidth]{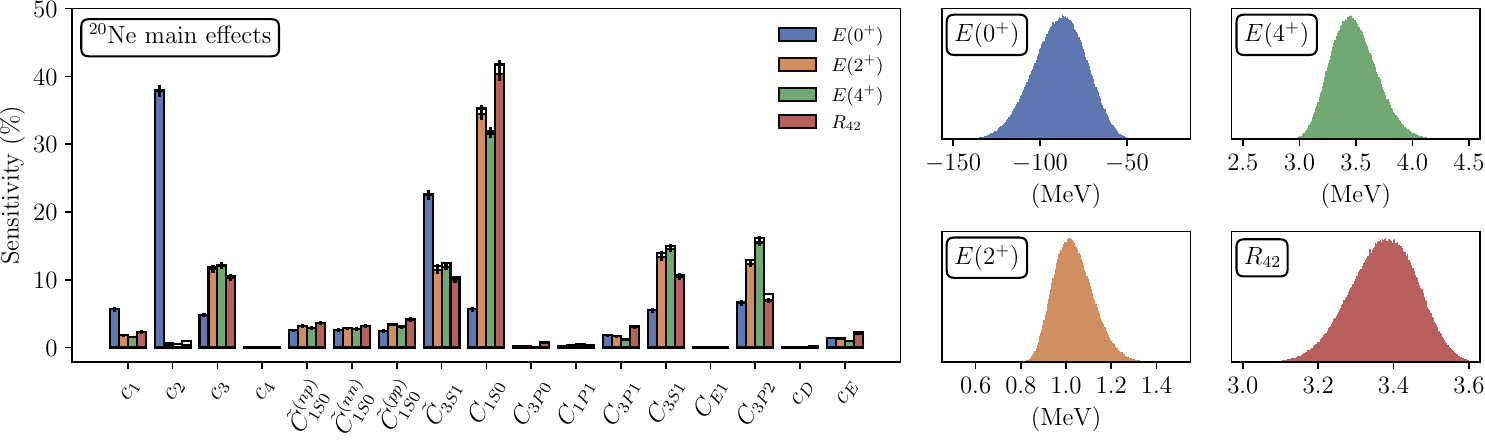}
    \includegraphics[width=0.9\linewidth]{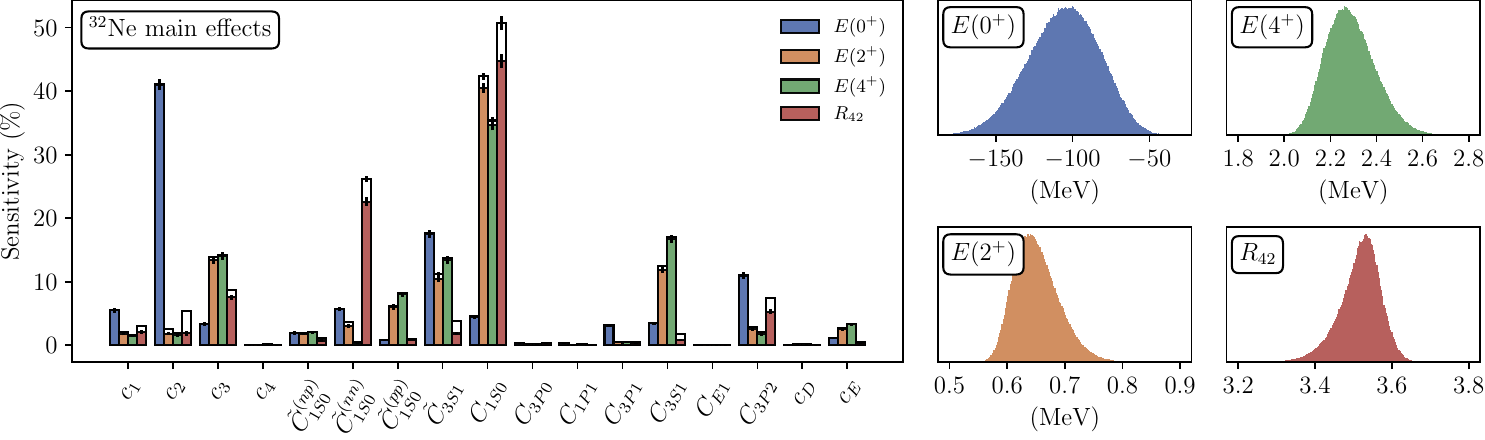}
    \includegraphics[width=0.9\linewidth]{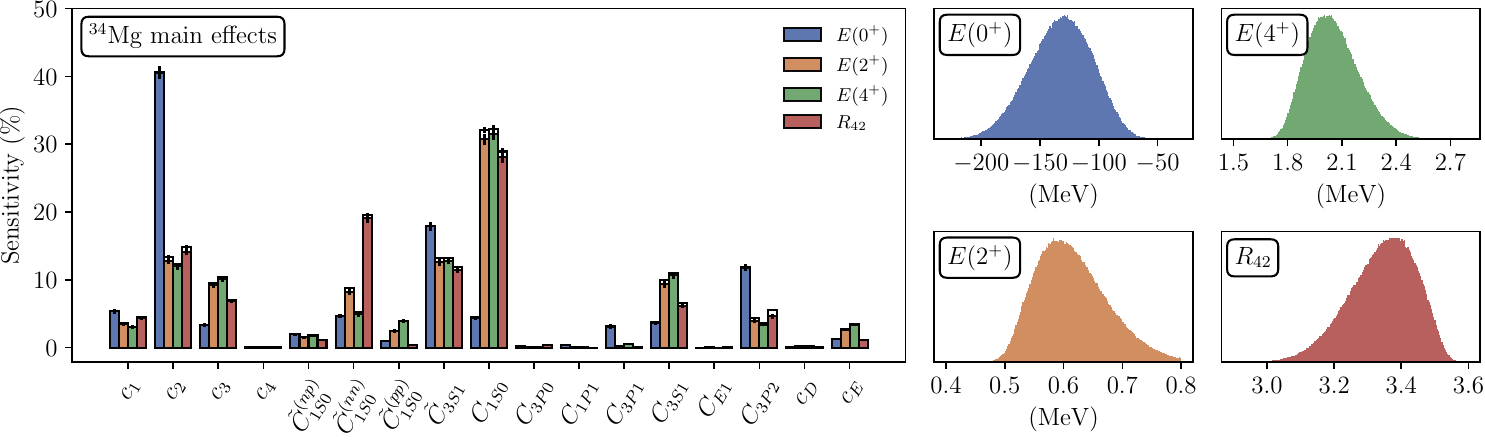}
    \caption{Results of the global sensitivity-analysis for the structure of $^{20,32}$Ne and $^{34}$Mg (top, center, and bottom panels). All Monte Carlo samples obtained as in the main text. Left panels: the mains effects (colored bars) for $R_{42}$ ground state energy ($E_{gs}$) and excited $2^+$ and $4^+$ states. The vertical black bars indicate the respective 95\% confidence intervals of the sensitivity indices as obtained using bootstrapped sampling. Here we also include total effects (white bars on top). Groups of right four panels: histograms displaying the variation of the 1,179,648 samples for each output. All energies were obtained using an emulator based on eigenvector continuation of projected Hartree-Fock.}
    \label{fig:all_gsa}
\end{figure*}

Figure~\ref{fig:all_gsa} shows the results from the sensitivity analysis of the energies $E(0^+)$, $E(2^+)$, and $E(4^+)$ in $^{20,32}$Ne and $^{34}$Mg. There are three main trends to observe. First, the variance of $E(0^+)$, i.e., the energy of the ground state, is explained to a great extent by the subleading pion-nucleon coupling $c_2$ and the leading $S-$wave contact $\tilde{C}_{^3S_1}$ in all three nuclei. The latter coupling is directly proportional to the deuteron binding energy. Second, the ground and excited state energies exhibit different main effect patterns and this indicates that the structure of their respective wave functions likely differ. Third, for the energies, we also show the total effects~\cite{SALTELLI2010259} (white bars on top of colored bars of the main effects). They are nearly identical to the main effects, with some differences observed in $^{32}$Ne, and this indicates that the (sum of) higher-order sensitivities are very small in the present domain. 

\begin{figure*}[hb!]
    \includegraphics[width=0.99\textwidth]{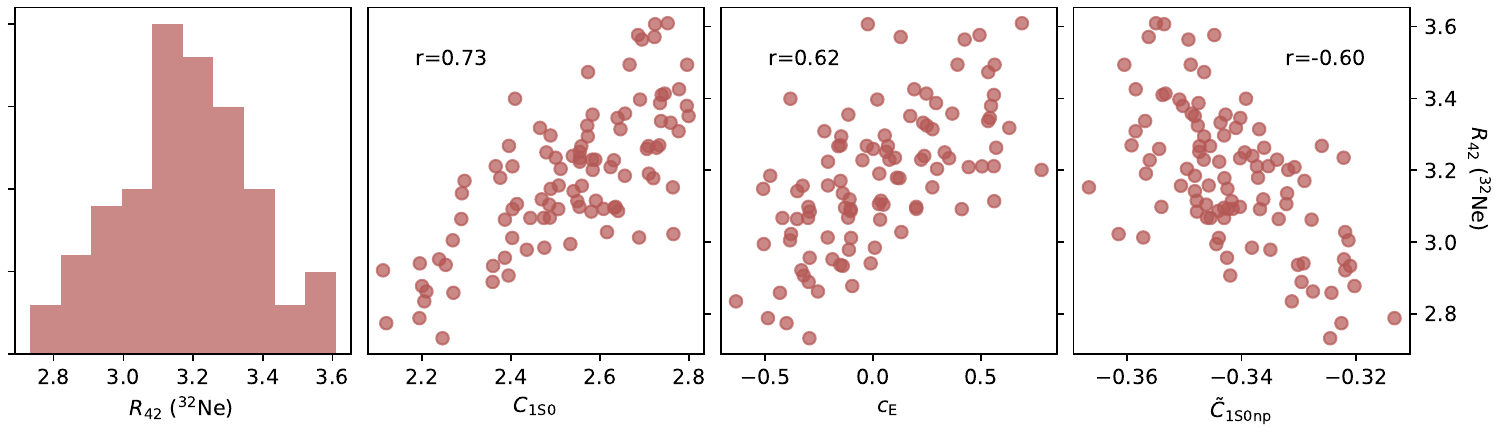}
    \includegraphics[width=0.99\textwidth]{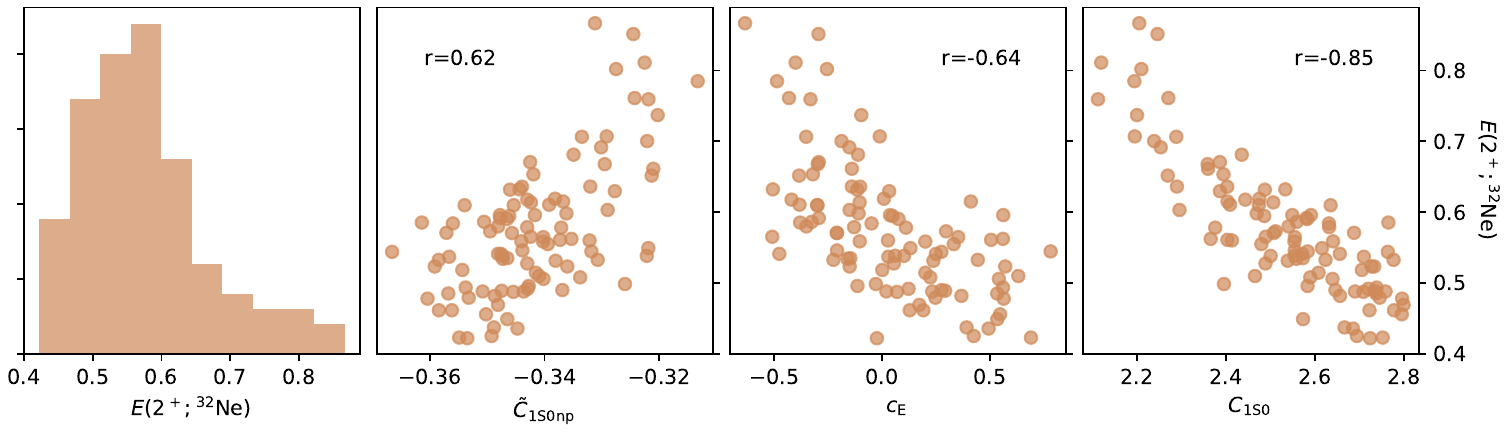}
    \caption{Correlation between the low-energy constants of the delta-full next-to-next-to leading order Hamiltonian and the observables $R_{42}$ (upper row), $E(2+)$ (lower row) of $^{32}$Ne. The scatter plots include the 100 interaction samples from the Bayesian posterior while the histograms show the distribution for the respective observable. The Pearson correlation coefficient, $r$, is extracted for the samples for each pair of variables. All other low-energy constants have correlation coefficients with an absolute magnitude smaller than 0.4 and are not shown.}
    \label{fig:ne20-R42_pars}
\end{figure*}

Figure~\ref{fig:ne20-R42_pars} shows the three strongest correlations between the low-energy constants of the delta-full NNLO Hamiltonian and the observables $R_{42}$ and  $E(2^+)$ of $^{32}$Ne  for the ensemble of Bayesian posterior interactions. We remind the reader that $c_E$ is the low-energy constant of the three-body contact, while $\tilde{C}_{\rm 1S0np}$ and $C_{\rm 1S0}$ are low-energy cofficients at (isospin-breaking) leading and next-to-leading order in the ${^1S_0}$ partial wave (the former acts in the neutron-proton channel of this partial wave). 

\clearpage
%


%

\end{document}